\documentclass[a4paper,11pt]{article}

\usepackage[latin1]{inputenc} %definit la table de caractère latine
\usepackage[english]{babel}
\usepackage{amsmath,amssymb,amsbsy,latexsym,textcomp}
\usepackage{bm}
\usepackage[dvips,xdvi]{graphicx}
\usepackage{pstricks}
\usepackage{subfig}

\newtheorem{theorem}{Theorem}%[section]
\newtheorem{remark}{Remark}%[section]
\newtheorem{proposition}{Proposition}%[section]
\newtheorem{lemma}{Lemma}%[section]
\newtheorem{assumption}{Assumption}%[section]

\newcommand{\rank}[1]{%
{\textup{rank}}{#1}}

\newcommand{\vect}[1]{%
{\textup{vec}}{#1}}

\newcommand{\col}[1]{%
{\textup{col}}{#1}}

\newcommand{\tr}[1]{%
{\textup{tr}}{#1}}

\newcommand{\range}[1]{%
{\textup{range}}{#1}}

\newcommand{\HRule}{\rule{\linewidth}{0.5mm}}

\begin{document}
\begin{titlepage}
\begin{center}

\includegraphics[width=0.3\textwidth]{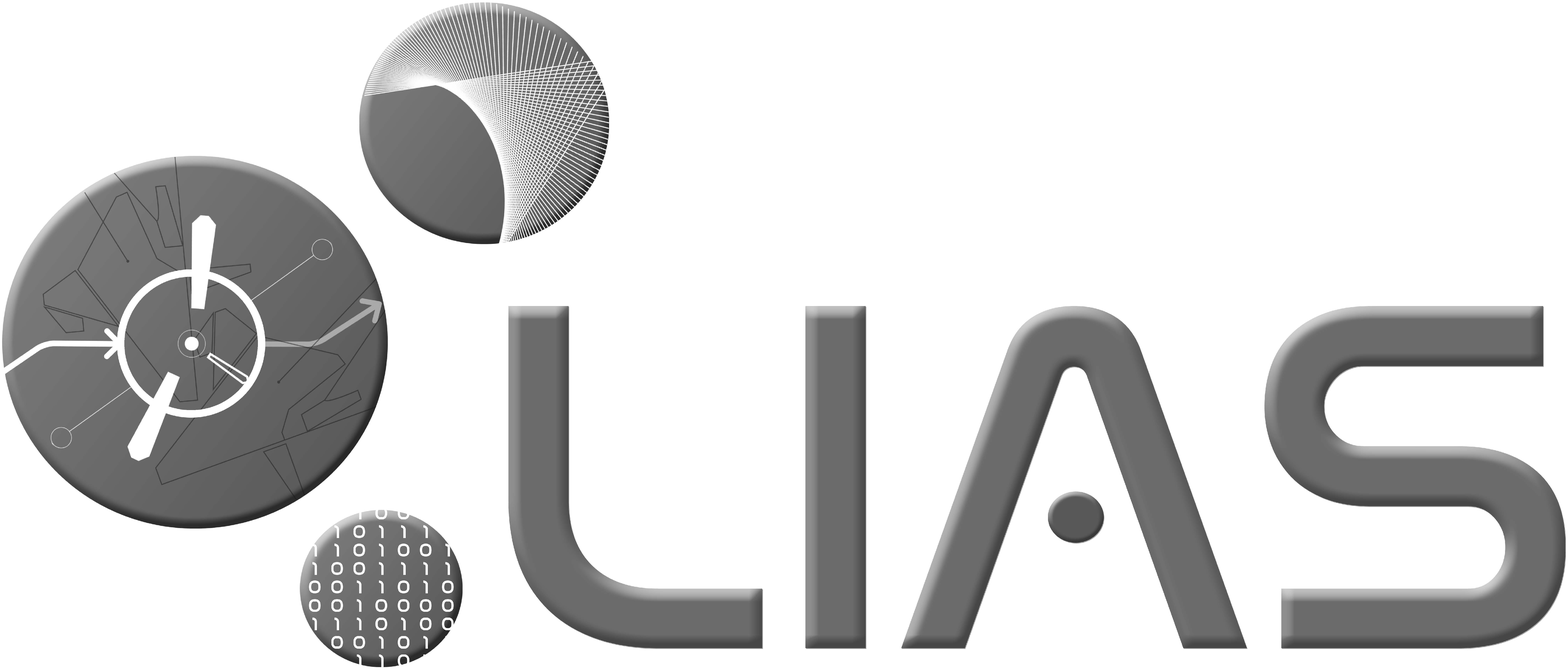} \\
\textsf{LIAS laboratory -- Poitiers University} % According to your institution
% \textsf{LIAS laboratory -- ENSMA}

\bigskip~\bigskip

\textsc{Technical Report\footnote{Technical reports from the Automatic Control group of the LIAS are available from \texttt{http://www.lias-lab.fr/publications/fulllist}}}

\bigskip

\HRule \\ \vspace{0.4cm}
\Large
\textsc{Regression techniques for subspace-based black-box state-space system identification: an overview} % Title

\HRule

\bigskip

\normalsize
\textsf{Author:} \\
\textsc{Guillaume Mercère\footnote{LIAS Automatic Control division}} 

\bigskip

\textsf{Email :} \\ \texttt{guillaume.mercere@univ-poitiers.fr}

\bigskip

\textsc{Report no: UP\_AS\_001}
\end{center}

\vfill
\begin{flushleft}
\textsf{Address :} \\
\textsf{Bâtiment B25 \\
2ème étage \\
2 rue Pierre Brousse \\
B.P. 633 \\
86022 Poitiers Cedex \\
web-site:} \texttt{http://www.lias-lab.fr/}
\end{flushleft}

\vfill
\begin{flushright}
\textsf{30 May 2013}
\end{flushright}

\end{titlepage}

\newpage
\thispagestyle{empty}
\mbox{}

\newpage
\setcounter{page}{1}

\title{\textsc{Regression techniques for subspace-based black-box state-space system identification: an overview\footnote{This document is available on \texttt{http://www.lias-lab.fr/perso/guillaumemercere/}}}}

\author{Guillaume Mercère\thanks{G. Mercère  is with the University of Poitiers, Laboratoire d'Informatique et d'Automatique pour les Systèmes, 40 avenue du Recteur Pineau, 86000 Poitiers, France. \texttt{guillaume.mercere@univ-poitiers.fr}}}

\date{\today}

\maketitle

\begin{abstract} % Abstract of not more than 200 words.
As far as the identification of linear time-invariant state-space representation is concerned, among all of the solutions available in the literature, the subspace-based state-space model identification techniques have proved their efficiency in many practical cases since the beginning of the 90's as illustrated, \emph{e.g.}, in \cite{VVG94,Bal97,Bal98,FMV00,JSL01,Cal02,Oku03,CMM06,WV09,BL11}. This paper introduces an overview of these techniques by focusing on their formulation as a least-squares problem. Apart from the article \cite{Qin06}, to the author's knowledge, such a regression formulation is not totally investigated in the books \cite{OM96,Kat05,VV07} which can be considered as the references as far as subspace-based identification is concerned. Thus, in this paper, a specific attention is payed to the regression-based techniques used to identify systems working under open-loop as well as closed-loop conditions.
\end{abstract}


%%%%%%%%%%%%%%%%%%%%%%%%%%%%%%%%%%%%%%%%%%%%%%%%%%%%%%%%%%%%%%%%%
%%%%%%%%%%%%%%%%%%%%%%%%%%%%% SECTION %%%%%%%%%%%%%%%%%%%%%%%%%%%
\section{Motivations and problem formulation}\label{para22:motivationsls4sid}

% Introduce PWA and work with Laurent about state-space hybrid system identification

% Introduce uncertainty domain.

% When the ARX or ARMAX model linked to the propagator is introduced, make an equivalent study as the one proposed in WJ94 for ARX models. 

% Introduce the link between the propagator and the Echelon form (see Peternell's PhD)

A linear time-invariant (LTI) discrete-time (DT) system can always be represented with the help of an LTI DT state-space form
\begin{subequations}\label{eq22:ssprocessform}
\begin{align}
\mathbf{x}(t+1) &= \mathbf{A} \mathbf{x}(t) + \mathbf{B} \mathbf{u}(t) + \mathbf{w}(t) \\
\mathbf{y}(t) &= \mathbf{C} \mathbf{x}(t) + \mathbf{D} \mathbf{u}(t) + \mathbf{v}(t)
\end{align}
\end{subequations}
where $\mathbf{x}(t) \in \mathbb{R}^{n_x}$ is the state vector, $\mathbf{u}(t) \in \mathbb{R}^{n_u}$ is the input vector, $\mathbf{y}(t) \in \mathbb{R}^{n_y}$ is the output vector, $\mathbf{v}(t) \in \mathbb{R}^{n_y}$ is the output measurement noise vector and $\mathbf{w}(t) \in \mathbb{R}^{n_x}$ is the process noise vector. $\left( \mathbf{A}, \mathbf{B}, \mathbf{C}, \mathbf{D} \right)$ are the state-space matrices of the system with appropriate dimensions.

\begin{remark}
In this paper, all the developments involve only DT data and DT systems. Thus, $t \in \mathbb{Z}$ and can be related to the sampling period $T_s$ used to acquire the input-output data. Notice however that one of the subspace-based identification algorithms introduced in the following overview has been recently extended in the continuous-time (CT) framework \cite{BL10,BL10b,BL11}.
\end{remark}

In order to estimate the state-space matrices $\left( \mathbf{A}, \mathbf{B}, \mathbf{C}, \mathbf{D} \right)$, the standard maximum likelihood methods \cite{Lju99} can be used. However, when the user wants to avoid a difficult non-linear optimization, an interesting alternative consists in resorting to a subspace-based state-space identification (4SID) algorithm \cite{OM96,Kat05,VV07}. The subspace-based identification methods have attracted a large interest since the 1990's because, among other things,
\begin{itemize}
\item they use reliable and robust numerical tools such as the RQ factorization or the singular values decomposition (SVD) \cite{GL96},
\item they do not require any non-linear optimization scheme,
\item they lead to fully-parameterized and well-conditioned black-box state-space forms (for instance a balanced realization \cite{McK94}),
\item they can handle multi-input-multi-output (MIMO) systems as easily as single-input-single-output (SISO) systems.
\end{itemize}
Historically, the first studies \cite{VD92,VD92b,Ver93,Ver94,OM94b,OM95,Vib95} mainly focused on the development of efficient algorithms which yield consistent estimates under different practical conditions and noise properties. This work has led to two main classes of subspace-based identification methods:
\begin{itemize}
\item the techniques which retrieve the space generated by the columns of the extended observability matrix from the available input-output data (\emph{e.g.}, the ``Multivariable Output-Error State sPace'' (MOESP) methods \cite{VD92,VD92b,Ver93,Ver94} or the ``Instrumental Variable for Subspace State-Space Sytem IDentification'' (IV-4SID) methods \cite{Vib95}),
\item the algorithms which estimate the state sequence  from the available input-output data (\emph{e.g.}, the ``Numerical algorithms for Subspace State-Space SyStem IDentification'' (N4SID) \cite{OM94b} or the ``Canonical Variate Analysis'' (CVA) method \cite{Lar90}.
\end{itemize}
It is interesting to point out that these methods
\begin{itemize}
\item share the same linear algebra tools (RQ factorization and SVD),
\item require projection techniques in order to calculate specific subspaces of the system and to fix the state coordinate basis \cite{Pet95} giving rise to a particular fully-parameterized state-space form, \emph{e.g.}, a balanced realization \cite{CM97,Mal05} or an orthogonal basis for the state-space \cite{OM95b},
\item involve model reduction techniques in order to determine the order of the system (especially when the data is corrupted by disturbances).
\end{itemize}
Interesting from a numerical point of view, these techniques
\begin{itemize}
\item are a lot more difficult to study from a statistical point of view than, \emph{e.g.}, the maximum likelihood techniques \cite{Lju99},
\item make the introduction of prior information about the system into the identification procedure quite complex.
\end{itemize}
These shortcomings mainly happen because the aforementioned subspace-based identification methods do not minimize explicitly a cost function in order to estimate the state-space matrices. Thus,
\begin{itemize}
\item the standard tools dedicated to the statistical analysis of the regression or prediction-based estimates \cite{Lju99} are difficult to adapt to this specific framework,
\item the techniques incorporating prior knowledge about the process into the optimization problem with the help of constrained equalities or inequalities \cite{Rot73,LH87} or via specific probabilistic tools \cite{Pet81} cannot be used directly with most of the subspace-based identification methods.
\end{itemize}
In order to adapt the tools used for the statistical analysis of the standard linear regression estimates to the subspace-based identification framework, several researchers have investigated the formulation of the subspace-based identification with the help of specific structured regression models \cite{PSD96,JW96,BJ00}. Initially developed for systems working under open-loop conditions \cite{LM96b,QL03b,QLL05}, this linear regression approach has allowed the development as well as the analysis of some subspace-based identification algorithms for closed-loop data \cite{LM96,Jan03,QL03,Jan05,LQL05,CP05,CP05b,HWV09b,Chi10}. More recently, this formulation of the subspace-based identification as a linear regression-based problem has been used to incorporate prior information into some subspace-based identification algorithms \cite{TH07,TH09,Aal10}.

The linear regression formulation of the subspace-based identification is introduced thoroughly in this paper. Open-loop as well as closed-loop operating conditions are respectively considered in Sub-Section~\ref{para22:ls4sidopenloop} and Sub-Section~\ref{para22:ls4sidclosedloop}. To the author's knowledge, such an overview is not available in any book dedicated to subspace-based identification. Notice also that this least-squares interpretation of the subspace-based identification techniques is the basis of the developments introduced in \cite{MB11}.

 % This specific study can be justified by noticing that
% \begin{itemize}
% \item the least-squares interpretation of the subspace-based identification 
% \begin{itemize}
% \item makes its understanding and its analysis much easier,
% \item helps the user to incorporate prior information into the subspace-based identification procedure,
% \end{itemize} 
% \item the parameterized state-space form introduced in Section~?? (see Chapter~??) and related to the propagator \cite{MD91,MBL08}
% \begin{itemize}
% \item can be obtained from the input-output data by using specific least-squares algorithms,
% \item can be linked to the high order ARX or ARMAX models put foward by \citeauthor{Jan03} or \citeauthor{Chi07} in \cite{Jan03,Chi07} (see also \cite{HWV09b} for a recent contribution).
% \end{itemize}
% \end{itemize}
% More precisely, the following... TO BE COMPLETED.

Before presenting the basic idea of the regression-based subspace identification, it is important to introduce generic notations which will be used all along this paper. First, for any vector $\mathbf{r}(t) \in \mathbb{R}^{n_r}$, we can define
\begin{itemize}
\item the infinite past stacked vector
\begin{equation}\label{eq22:geninfpastvec}
\mathbf{r}^-(t) =
\begin{bmatrix}
\vdots \\
\mathbf{r}(t-2) \\
\mathbf{r}(t-1)
\end{bmatrix}
\end{equation}
\item the infinite future stacked vector
\begin{equation}\label{eq22:geninffutvec}
\mathbf{r}^+(t) =
\begin{bmatrix}
\mathbf{r}(t) \\
\mathbf{r}(t+1) \\
\vdots
\end{bmatrix}
\end{equation}
\item the finite past stacked vector
\begin{equation}\label{eq22:genfinpastvec}
\mathbf{r}_\ell^-(t) =
\begin{bmatrix}
\mathbf{r}(t-\ell) \\
\vdots \\
\mathbf{r}(t-2) \\
\mathbf{r}(t-1)
\end{bmatrix} \in \mathbb{R}^{\ell n_r \times 1}, \ \ell \in \mathbb{N}_*^+
\end{equation}
\item the finite future stacked vector
\begin{equation}\label{eq22:genfinfutvec}
\mathbf{r}_\ell^+(t) =
\begin{bmatrix}
\mathbf{r}(t) \\
\mathbf{r}(t+1) \\
\vdots \\
\mathbf{r}(t+\ell-1)
\end{bmatrix} \in \mathbb{R}^{\ell n_r \times 1}, \ \ell \in \mathbb{N}_*^+ .
\end{equation}
\end{itemize}
By having access to these finite stacked vectors\footnote{By looking closer at the definition of $\mathbf{r}_\ell^-(t)$ and $\mathbf{r}_\ell^+(t)$, it is quite obvious that a single definition could be used because, \emph{e.g.}, $\mathbf{r}_\ell^-(t) = \mathbf{r}_\ell^+(t-\ell)$. However, in order to highlight the fact that $\mathbf{r}_\ell^-(t)$ (resp. $\mathbf{r}_\ell^+(t)$) is composed of past (resp. future) data with respect to the present time index $t$, two notations are used hereafter. This notion of past and future is indeed standard when subspace-based identification is concerned.}, the past and future Hankel matrices (resp. $\mathbf{R}_{\ell,M}^-(t)$ and $\mathbf{R}_{\ell,M}^+(t)$) can be deduced as follows
\begin{multline}\label{eq22:pasthankelmat}
 \mathbf{R}_{\ell,M}^-(t) =
\begin{bmatrix} 
\mathbf{r}_\ell^-(t) & \cdots & \mathbf{r}_\ell^-(t+M-1)
\end{bmatrix} \in \mathbb{R}^{\ell n_r \times M}, \ell \in \mathbb{N}_*^+ ,\ M \in \mathbb{N}_*^+
\end{multline}
\begin{multline}\label{eq22:futhankelmat}
 \mathbf{R}_{\ell,M}^+(t) =
\begin{bmatrix} 
\mathbf{r}_\ell^+(t) & \cdots & \mathbf{r}_\ell^+(t+M-1)
\end{bmatrix} \in \mathbb{R}^{\ell n_r \times M}, \ell \in \mathbb{N}_*^+ ,\ M \in \mathbb{N}_*^+ .
\end{multline}
Now, from four matrices $\mathbf{A}$, $\mathbf{B}$, $\mathbf{C}$ and $\mathbf{D}$ of appropriate dimensions, the following generic block matrices can be constructed
\begin{equation}\label{eq22:contmat}
\bm{\Omega}_\ell(\mathbf{A},\mathbf{B}) =
\begin{bmatrix}
\mathbf{A}^{\ell-1} \mathbf{B} & \cdots & \mathbf{A} \mathbf{B} & \mathbf{B} 
\end{bmatrix}, \ \ell \in \mathbb{N}_*^+
\end{equation}
\begin{equation}\label{eq22:obsmat}
  \bm{ \Gamma}_\ell(\mathbf{A},\mathbf{B}) =
  \begin{bmatrix}
    \mathbf{B} \\
\mathbf{B} \mathbf{A} \\
\vdots \\
\mathbf{B} \mathbf{A}^{\ell-1}
  \end{bmatrix}, \ \ell \in \mathbb{N}_*^+
\end{equation}
\begin{equation}\label{eq22:gentoeplitz}
  \mathbf{H}_\ell(\mathbf{A},\mathbf{B},\mathbf{C},\mathbf{D}) =
\begin{bmatrix} 
\mathbf{D}& \mathbf{0} & \cdots & \mathbf{0}\\
\mathbf{C}\mathbf{B} & \mathbf{D} & \cdots &\mathbf{0}\\
\vdots & \ddots & \ddots &\vdots\\
\mathbf{C}\mathbf{A}^{\ell-2}\mathbf{B} & \cdots & \mathbf{C}\mathbf{B} & \mathbf{D}
\end{bmatrix}, \ \ell \in \mathbb{N}_*^+ .
\end{equation}
By construction, for $\ell \geq n_x$, $\bm{ \Gamma}_\ell(\mathbf{A},\mathbf{C})$ (resp. $\bm{\Omega}_\ell(\mathbf{A},\mathbf{B})$) is the extended observability (resp. (reversed) controllability) matrix of the system~\eqref{eq22:ssprocessform}.

%%%%%%%%%%%%%%%%%%%%%%%%%%%%%%%%%%%%%%%%%%%%%%%%%%%%%%%%%%%%%%%%%
%%%%%%%%%%%%%%%%%%%%%%%%%%%%% SECTION %%%%%%%%%%%%%%%%%%%%%%%%%%%
\section{Subspace-based identification involving linear regression}\label{para22:ls4sidreview}

A quick look at the literature dedicated to the linear regression-based subspace identification \cite{WJ94,JW96,LM96,PSD96,Bau03,Qin06} indicates that most of the least-squares methods for subspace-based identification first try to determine a sequence of the state vector $\mathbf{x}$. By having access to a reliable state estimate as well as a sufficiently large set of input and output measurements, the identification problem becomes a linear regression problem
\begin{equation}
\bm{\varpi}(t) = \bm{\Theta} \bm{\phi}(t) + \bm{\nu}(t)
\end{equation}
where
\begin{align}
\bm{\varpi}(t) &=
\begin{bmatrix}
\mathbf{x}(t+1) \\
\mathbf{y}(t)
\end{bmatrix} &
\bm{\Theta} =&
\begin{bmatrix}
\mathbf{A} & \mathbf{B} \\
\mathbf{C} & \mathbf{D}
\end{bmatrix} &
\bm{\phi}(t) &=
\begin{bmatrix}
\mathbf{x}(t) \\
\mathbf{u}(t)
\end{bmatrix} &
\bm{\nu}(t) &=
\begin{bmatrix}
\mathbf{v}(t) \\
\mathbf{w}(t)
\end{bmatrix} .
\end{align}
Second, the parameter matrix $\bm{\Theta}$ can be estimated by using a least-squares method. The characteristics of the disturbances $\mathbf{v}$ and $\mathbf{w}$ can also be estimated, in a second step, from the analysis of the residuals. 

\begin{remark}
As soon as the state vector is determined, the coordinate basis of the state-space realization is fixed.
\end{remark}

In this section, the main steps involved in this specific linear regression framework are described in a quite general context. Sub-Section~\ref{para22:stateconstruct} is more precisely dedicated to the construction of a state sequence from the available input and output data. By knowing this state sequence (as a linear combination of past data), specific multi-step predictors are introduced in Sub-Section~\ref{para22:ls4sidopenloop} as well as dedicated optimization algorithms (involving certain rank constraints) in order to yield consistent estimates of the state-space matrices $\mathbf{A}, \mathbf{B}, \mathbf{C}, \mathbf{D}$. Because the linear regression techniques studied in Sub-Section~\ref{para22:ls4sidopenloop} lead to biased estimates when the system to identify operates in closed-loop, a particular attention is paid to techniques efficient under closed-loop operating conditions in Sub-Section~\ref{para22:ls4sidclosedloop}.

%%%%%%%%%%%%%%%%%%%%%%%%%%% SUBSECTION %%%%%%%%%%%%%%%%%%%%%%%%%%
\subsection{Construction of a state sequence from past input and past output data}\label{para22:stateconstruct}

This sub-section is mainly dedicated to this first step of the main linear regression-based subspace identification methods. Different formulations of the state sequence as specific combinations of past inputs and past outputs are more precisely addressed. Notice right now that these combinations will involve unknown matrices such as the state-space matrices of the system (via, \emph{e.g.}, the Markov parameters). These matrices are unknown at this step of the identification procedure. The goal of this section is not to estimate the state but to reformulate it as a combination of known signals in order to simplify the data equations used in Sub-Section~\ref{para22:ls4sidopenloop} and Sub-Section~\ref{para22:ls4sidclosedloop}. The problem of the state-space matrices extraction is indeed investigated in these two aforementioned paragraphs. This state sequence determination phase is only an intermediate step necessary to give rise to linear regression problems.

% SUBSUBSECTION %
\subsubsection{State sequence construction from the Kalman filter}

A standard way to construct a state sequence consists in resorting to an observer. An observer is indeed a mathematical tool able to approximate the state vector of a dynamic system from measurements of its inputs and outputs. A standard observer is the Kalman filter. The Kalman filter is a well-known and widely used tool (see, \emph{e.g.}, \cite{FP09,GT10,GA10} for recent applications). One of its main use consists in reconstructing the state of a given state-space system in a statistically optimal way \cite{Kal60,KB61}. By statistically optimal way, it is meant that the Kalman filter yields an unbiased state estimate $\hat{\mathbf{x}}(t)$ (\emph{i.e.} $\mathbb{E}{\left\{ \mathbf{x}(t) - \hat{\mathbf{x}}(t) \right\}} = \mathbf{0}$) with\footnote{$\mathbb{E}{\left\{ \bullet \right\}}$ stands for the expected value of the random variable $\bullet$.} a state error covariance matrix $\mathbb{E}{\left\{ \left( \mathbf{x}(t) - \hat{\mathbf{x}}(t) \right) \left( \mathbf{x}(t) - \hat{\mathbf{x}}(t) \right)^\top \right\}}$ as small as possible \cite{Joh93,Gus00,VV07}. % Thus, the Kalman filter is the best estimate of the state in the mean square sense \cite{Bau03}. 
Contrary to a standard Luenberger observer, the Kalman filter takes the disturbances acting on the system into account.

The Kalman filter deals with linear time-varying systems of the form
\begin{subequations}\label{eq22:tvssprocessform}
\begin{align}
\mathbf{x}(t+1) &= \mathbf{A}(t) \mathbf{x}(t) + \mathbf{B}(t) \mathbf{u}(t) + \mathbf{w}(t) \\
\mathbf{y}(t) &= \mathbf{C}(t) \mathbf{x}(t) + \mathbf{D}(t) \mathbf{u}(t) + \mathbf{v}(t)
\end{align}
\end{subequations}
where the process noise $\mathbf{w}$ and the measurement noise $\mathbf{v}$ are assumed to be zero-mean white noises with joint covariance matrix
\begin{equation}
\mathbb{E}\left\{
\begin{bmatrix}
\mathbf{v}(k) \\
\mathbf{w}(k)
\end{bmatrix}
\begin{bmatrix}
\mathbf{v}^\top(j) & \mathbf{w}^\top(j)
\end{bmatrix}
\right\} =
\begin{bmatrix}
\mathbf{R}(k) & \mathbf{S}^\top(k) \\
\mathbf{S}(k) & \mathbf{Q}(k)
\end{bmatrix}
\delta_{k j} \geq 0
\end{equation}
where $\mathbf{R}(k) > 0$, $\mathbf{Q}(k) \geq 0$ and where $\delta_{k j}$ is the Kronecker delta function defined by
\begin{equation}\label{eq22:kroneckerdelta}
  \delta_{k j} = 
\left\{ \begin{matrix}
1 & \text{ if } \  k = j \\
0 & \text{ if } \  k \neq j
\end{matrix} \right. .
\end{equation}
The Kalman filter is a recursive estimator. Most of the time, it is written as a two-step approach where \cite{WB01}
\begin{itemize}
\item the first step (usually named the prediction phase) uses the state estimate from the previous time-step in order to calculate an estimate of the state at the current time-step. This predicted state estimate is also known as the \emph{a priori} state estimate because it does not include observation information from the current time-step.
\item the second step (usually named the update phase) combines the current \emph{a priori} prediction with current observation information in order to refine the state estimate. This update state estimate is also named the \emph{a posteriori} state estimate.
\end{itemize}
The DT Kalman filter can be viewed as a cycle where the prediction step and the correction step alternate.

In the following, only the prediction phase is introduced. For more details concerning the DT Kalman filter, see, \emph{e.g.}, \cite{AM90}. This limitation can be justified by the fact that, in the system identification framework considered herein, at best the data up to time $t-1$ are involved to estimate the unknown parameters. Thus, the update phase will not intervene in the following.

The prediction step can be described as follows. By assuming that an estimate $\hat{\mathbf{x}}$ of $\mathbf{x}$ is available at time $t$ satisfying
\begin{subequations}
\begin{align}
\mathbb{E}\left\{ \mathbf{x}(t) \right\} &= \mathbb{E}\left\{ \hat{\mathbf{x}}(t) \right\} \\
\mathbb{E}\left\{ \left( \mathbf{x}(t) - \hat{\mathbf{x}}(t) \right) \left( \mathbf{x}(t) - \hat{\mathbf{x}}(t) \right)^\top \right\} &= \mathbf{P}(t) \geq 0 ,
\end{align}
\end{subequations}
the state estimate at time $t+1$ is given by \cite[Chapter 5]{Kat05}
\begin{equation}\label{eq22:kalmanstateupdate}
\hat{\mathbf{x}}(t+1) = \mathbf{A}(t) \hat{\mathbf{x}}(t) + \mathbf{B}(t) \mathbf{u}(t) + \bm{\mathcal{K}}(t) \left( \mathbf{y}(t) - \mathbf{C}(t) \hat{\mathbf{x}}(t) - \mathbf{D}(t) \mathbf{u}(t) \right)
\end{equation}
where the Kalman gain $\bm{\mathcal{K}}$ satisfies
\begin{equation}\label{eq22:kalmangainupdate}
\bm{\mathcal{K}}(t) = \left( \mathbf{S}(t) + \mathbf{A}(t) \mathbf{P}(t) \mathbf{C}^\top(t) \right) \left( \mathbf{R}(t) + \mathbf{C}(t) \mathbf{P}(t) \mathbf{C}^\top(t) \right)^{-1}
\end{equation}
where the state error covariance matrix $\mathbf{P}(t)$ is updated by using the Riccati difference equation
\begin{multline}\label{eq22:riccatiupdate}
\mathbf{P}(t+1) = \mathbf{A}(t) \mathbf{P}(t) \mathbf{A}^\top(t) + \mathbf{Q}(t) - \left( \mathbf{S}(t) + \mathbf{A}(t) \mathbf{P}(t) \mathbf{C}^\top(t) \right) \\
\left( \mathbf{R}(t) + \mathbf{C}(t) \mathbf{P}(t) \mathbf{C}^\top(t) \right)^{-1} 
\left( \mathbf{S}(t) + \mathbf{A}(t) \mathbf{P}(t) \mathbf{C}^\top(t) \right)^{\top} .
\end{multline}
%These formulas can be seen as a concatenation of the two steps composing the conventional Kalman filter, \emph{i.e.}, the calculation of the filtered state followed by the computation of the one-step-ahead predicted state.
Furthermore, the conditional expectation of the output signal $\mathbf{y}(t)$, given the input and output data from the infinite past up to time $t-1$, satisfies
\begin{equation}
  \label{eq22:outputexp}
  \hat{\mathbf{y}}(t) = \mathbf{C}(t) \hat{\mathbf{x}}(t) + \mathbf{D}(t) \mathbf{u}(t) .
\end{equation}

The developments introduced up until now have handled time-varying systems. As shown in \cite[Chapter 5]{Kat05}, when the state-space matrices $\left( \mathbf{A}, \mathbf{B}, \mathbf{C}, \mathbf{D} \right)$ as well as the variance matrices  $\left( \mathbf{R}, \mathbf{S}, \mathbf{Q} \right)$ are time-invariant, Eq.~\eqref{eq22:kalmanstateupdate},~\eqref{eq22:kalmangainupdate},~\eqref{eq22:riccatiupdate} and~\eqref{eq22:outputexp} can be adapted for time-invariant systems (see also \cite[Chapter 5]{VV07} for details). More precisely, by considering that the following assumptions are satisfied,
\begin{assumption}\label{ass:noise}
The noise terms $\mathbf{v}$ and $\mathbf{w}$ in the LTI state-space representation~\eqref{eq22:ssprocessform} are independent zero-mean white Gaussian noises with finite covariance matrices, \emph{i.e.},
\begin{equation}
  \mathbb{E}{
\left\{
\begin{bmatrix}
\mathbf{v}(k) \\
\mathbf{w}(k)
\end{bmatrix}
\begin{bmatrix}
\mathbf{v}^\top(\ell) & \mathbf{w}^\top(\ell)
\end{bmatrix}
\right\}
} =
\begin{bmatrix}
\mathbf{R} & \mathbf{S}^\top \\
 \mathbf{S} &  \mathbf{Q}
\end{bmatrix} \delta_{k \ell}
\end{equation}
where $\delta_{k \ell}$ is the Kronecker delta function (see Eq.~\eqref{eq22:kroneckerdelta}).
\end{assumption}
\begin{assumption}\label{ass:minimal}
  The LTI state-space system~\eqref{eq22:ssprocessform} is minimal, \emph{i.e.}, the pair $(\mathbf{A}, \mathbf{C})$ is observable and the pair $(\mathbf{A}, [\mathbf{B}, \mathbf{Q}^{1/2}])$ is controllable.
\end{assumption}
\noindent the following theorem can be stated \cite{AM90}.
\begin{theorem}\label{th:ltikalmanfilter}
  Consider the LTI system~\eqref{eq22:ssprocessform} and assume that the hypotheses~\ref{ass:noise} and \ref{ass:minimal} are satisfied. Then, 
\begin{itemize}
\item the Kalman filter is expressed as
\begin{equation}
\hat{\mathbf{x}}(t+1) = \mathbf{A} \hat{\mathbf{x}}(t) + \mathbf{B} \mathbf{u}(t) + \bm{\mathcal{K}}(t) \left( \mathbf{y}(t) - \mathbf{C} \hat{\mathbf{x}}(t) - \mathbf{D} \mathbf{u}(t) \right) ,
\end{equation}
\item the error covariance matrix satisfies the Riccati equation
\begin{multline}
\mathbf{P}(t+1) = \mathbf{A} \mathbf{P}(t) \mathbf{A}^\top + \mathbf{Q} - \left( \mathbf{S} + \mathbf{A} \mathbf{P}(t) \mathbf{C}^\top \right) \\
\left( \mathbf{R} + \mathbf{C} \mathbf{P}(t) \mathbf{C}^\top \right)^{-1} 
\left( \mathbf{S} + \mathbf{A} \mathbf{P}(t) \mathbf{C}^\top \right)^{\top}  ,
\end{multline}
\item the Kalman gain matrix $\bm{\mathcal{K}}(t)$ satisfies
\begin{equation}
\bm{\mathcal{K}}(t) = \left( \mathbf{S} + \mathbf{A} \mathbf{P}(t) \mathbf{C}^\top \right) \left( \mathbf{R} + \mathbf{C} \mathbf{P}(t) \mathbf{C}^\top \right)^{-1} ,
\end{equation}
\item the conditional expectation of the output signal $\mathbf{y}(t)$ is expressed as
\begin{equation}
\hat{\mathbf{y}}(t) = \mathbf{C} \hat{\mathbf{x}}(t) + \mathbf{D} \mathbf{u}(t) .
\end{equation}
\end{itemize}
Furthermore, for any symmetric initial condition $\mathbf{P}(0) > 0$,
\begin{equation}
  \lim_{t \rightarrow \infty} \mathbf{P}(t) = \mathbf{P} > \mathbf{0}
\end{equation}
where $\mathbf{P}$ satisfies
\begin{multline}
\mathbf{P} = \mathbf{A} \mathbf{P} \mathbf{A}^\top + \mathbf{Q} - \left( \mathbf{S} + \mathbf{A} \mathbf{P} \mathbf{C}^\top \right) \\
\left( \mathbf{R} + \mathbf{C} \mathbf{P} \mathbf{C}^\top \right)^{-1} 
\left( \mathbf{S} + \mathbf{A} \mathbf{P} \mathbf{C}^\top \right)^{\top} .
\end{multline}
Moreover, this matrix $\mathbf{P}$ is unique and the deduced Kalman gain matrix $\bm{\mathcal{K}}$ defined as
\begin{equation}
\bm{\mathcal{K}} = \left( \mathbf{S} + \mathbf{A} \mathbf{P} \mathbf{C}^\top \right) \left( \mathbf{R} + \mathbf{C} \mathbf{P} \mathbf{C}^\top \right)^{-1}
\end{equation}
ensures that the matrix $\tilde{\mathbf{A}} = \mathbf{A} - \bm{\mathcal{K}} \mathbf{C}$ satisfies
  \begin{equation}
    \lambda_{\textup{max}} (\tilde{\mathbf{A}}) < 1
  \end{equation}
where $\lambda_{\textup{max}} (\mathbf{A})$ denotes the eigenvalue of $\mathbf{A}$ of maximum modulus.
\end{theorem}
In this time-invariant framework, the Kalman filter induces two important consequences as far as system identification is concerned. 
\begin{itemize}
\item First, by following the lines of \cite[Appendix A.5, pp. 207--210]{OM96}, the (non-steady-state) Kalman filter can be written as a linear combination of past inputs, past outputs and an initial state estimate through the following theorem (see also \cite{OM94b})
\begin{theorem}
Given $\hat{\mathbf{x}}(0)$, $\mathbf{P}(0)$, samples of the input and output signals on the time range $\left[0, t \right]$, \emph{i.e.}, $\left\{ \mathbf{u}(k) \right\}_{k=0}^t$ and  $\left\{ \mathbf{y}(k) \right\}_{k=0}^t$ and all the matrices $\left( \mathbf{A}, \mathbf{B}, \mathbf{C}, \mathbf{D} \right)$ as well as the variance matrices  $\left( \mathbf{R}, \mathbf{S}, \mathbf{Q} \right)$, then the Kalman filter state $\hat{\mathbf{x}}(t)$ can be written as
\begin{multline}\label{eq22:KFstateestimate}
\hat{\mathbf{x}}(t)=
% \begin{bmatrix}
\left( \mathbf{A}^t - \bm{\Delta}_t \bm{\Gamma}_t(\mathbf{A},\mathbf{C}) \right) \hat{\mathbf{x}}(0)+ \bm{\Delta}_t \mathbf{y}_t^-(t) \\
 + \left( \bm{\Omega}_t(\mathbf{A},\mathbf{B}) - \bm{\Delta}_t \mathbf{H}_t(\mathbf{A},\mathbf{B},\mathbf{C},\mathbf{D}) \right) \mathbf{u}_t^-(t)
% \end{bmatrix}
% \begin{bmatrix}
% \hat{\mathbf{x}}(0) \\
% %\hline
% \rule{1cm}{.1pt} \\
% \mathbf{u}(0) \\
% \vdots \\
% \mathbf{u}(t-1) \\
% %\hline
% \rule{1cm}{.1pt} \\
% \mathbf{y}(0) \\
% \vdots \\
% \mathbf{y}(t-1) \\
% \end{bmatrix}
\end{multline}
where $\bm{\Omega}_t (\mathbf{A},\mathbf{B})\in \mathbb{R}^{n_x \times n_u t}$ is the (reversed) extended controllability matrix of the system, where $\bm{\Gamma}_t(\mathbf{A},\mathbf{C}) \in \mathbb{R}^{n_y t \times n_x}$ is the extended observability matrix of the system, where $\mathbf{H}_t(\mathbf{A},\mathbf{B},\mathbf{C},\mathbf{D}) \in \mathbb{R}^{n_y t \times n_u t}$ is a Toeplitz matrix defined as in Eq.~\eqref{eq22:gentoeplitz} and where $\bm{\Delta}_t \in \mathbb{R}^{n_x \times n_y t}$ can be defined by the following recursive formula
\begin{equation}
\bm{\Delta}_t =
\begin{bmatrix}
(\mathbf{A} - \bm{\mathcal{K}}(t-1) \mathbf{C} ) \bm{\Delta}_{t-1} &  \bm{\mathcal{K}}(t-1)
\end{bmatrix}.
\end{equation}
\end{theorem}
The main interest of this theorem is that the state sequence can be related to the past input and output data (and the memory of the system dynamics $\hat{\mathbf{x}}(0)$) with the help of the Kalman filter. As shown hereafter, this relation will lead to a data equation (see Eq.~\eqref{eq22:vectordataequ} or \cite{Vib95} for a definition) gathering only input and output signals.
\item Second, by assuming that the conditions~\ref{ass:noise} and \ref{ass:minimal} are satisfied, the process state-space form~\eqref{eq22:ssprocessform} is equivalent to the innovation form
\begin{subequations}\label{eq22:ssinnovform}
\begin{align}
\mathbf{x}(t+1) &= \mathbf{A} \mathbf{x}(t) + \mathbf{B} \mathbf{u}(t) + \bm{\mathcal{K}} \mathbf{e}(t) \label{eq22:stateequinnovform} \\
\mathbf{y}(t) &= \mathbf{C} \mathbf{x}(t) + \mathbf{D} \mathbf{u}(t) + {\mathbf{e}(t)} \label{eq22:outputequinnovform}
\end{align}
\end{subequations}
where $\bm{\mathcal{K}}$ is the steady-state Kalman filter gain defined in Theorem~\ref{th:ltikalmanfilter} and $\mathbf{e}$ is called the innovation vector \cite{Lju99}. More precisely, the innovation vector $\mathbf{e}$ is the part of $\mathbf{y}$ which cannot be predicted from the past data, \emph{i.e.},
\begin{equation}
  \mathbf{e}(t) = \mathbf{y}(t) - \hat{\mathbf{y}}(t) = \mathbf{C} ( \mathbf{x}(t) - \hat{\mathbf{x}}(t) ) + \mathbf{v}(t) .
\end{equation}
By construction (see Theorem~\ref{th:ltikalmanfilter}), for any realization of $\mathbf{v}$ and $\mathbf{w}$ satisfying Assumption~\ref{ass:noise}, a unique matrix $\bm{\mathcal{K}}$ and a unique vector $\mathbf{e}$ exist such that the representations~\eqref{eq22:ssprocessform} and~\eqref{eq22:ssinnovform} have the same input-output behavior \cite{Hav01}. Notice also that, although the Kalman gain is computed from the matrices $\left( \mathbf{A}, \mathbf{B}, \mathbf{C}, \mathbf{D} \right)$ as well as the variance matrices  $\left( \mathbf{R}, \mathbf{S}, \mathbf{Q} \right)$ in a complicated manner (see Theorem~\ref{th:ltikalmanfilter}), such an innovation form (deduced from Eq.~\eqref{eq22:ssprocessform}) verifies that all the eigenvalues of $\tilde{\mathbf{A}} = \mathbf{A} - \bm{\mathcal{K}} \mathbf{C}$ are strictly inside the unit circle when Assumptions~\ref{ass:noise} and \ref{ass:minimal} are satisfied. Indeed \cite{Kai80},
\begin{lemma}
  Given matrices $\mathbf{A} \in \mathbb{R}^{n_x \times n_x}$ and $\mathbf{C} \in \mathbb{R}^{n_y \times n_x}$, if the pair $(\mathbf{A},\mathbf{C})$ is observable, then a matrix $\bm{\mathcal{K}} \in \mathbb{R}^{n_x \times n_y}$ exists such that $\mathbf{A} - \bm{\mathcal{K}} \mathbf{C}$ is asymptotically stable.
\end{lemma}
This property means that the system is strict minimum phase because \cite{Pet95}
\begin{proposition}\label{prop:miniphase}
  A stable system represented by the LTI state-space form~\eqref{eq22:ssinnovform} is strict minimum phase if and only if
  \begin{equation}
    \lambda_{\textup{max}} (\mathbf{A} - \bm{\mathcal{K}} \mathbf{C}) < 1 .
  \end{equation}
\end{proposition}
As shown hereafter, this minimum phase property is really important in order to simplify some equations and is, in a way, the keystone for the recent closed-loop subspace-based identification techniques \cite{Jan03,CP05} (see Sub-Section~\ref{para22:ls4sidclosedloop} for details).
\end{itemize}

% SUBSUBSECTION %
\subsubsection{State sequence construction from infinite time series}

In the previous paragraph, the state sequence (see Eq.~\eqref{eq22:KFstateestimate}) has been written as a combination of finite input and output data sets plus a term involving the initial state estimate $\hat{\mathbf{x}}(0)$. This specificity implies that different Kalman filter sequences can be obtained according to the choice or the estimate of the initial state $\hat{\mathbf{x}}(0)$. This characteristic of the Kalman filter estimate~\eqref{eq22:KFstateestimate} has given rise to different state-space subspace-based identification algorithms according to the way this initial state sequence is chosen by the user (see, \emph{e.g.}, \cite[Algorithms 1 and 2 of Chapter 4]{OM96} for two major illustrations). Instead of resorting to the memory of the system dynamics $\hat{\mathbf{x}}(0)$, several authors, \emph{e.g.}, {K. Peternell} \cite{Pet95} or {D. Bauer} \cite{Bau98}, have chosen to tackle the problem of the state sequence construction by considering infinite input and output time series. The starting point of this approach is the state-space predictor form defined as \cite{Qin06} 
\begin{subequations}\label{eq22:sspredform}
\begin{align}
\mathbf{x}(t+1) &= \underbrace{(\mathbf{A} - \bm{\mathcal{K}} \mathbf{C})}_{\tilde{\mathbf{A}}} \mathbf{x}(t) + \underbrace{(\mathbf{B} - \bm{\mathcal{K}} \mathbf{D})}_{\tilde{\mathbf{B}}} \mathbf{u}(t) + \bm{\mathcal{K}} \mathbf{y}(t) \label{eq22:stateequpredform} \\
\mathbf{y}(t) &= \mathbf{C} \mathbf{x}(t) + \mathbf{D} \mathbf{u}(t) + {\mathbf{e}(t)} \label{eq22:outputequpredform}
\end{align}
\end{subequations}
which is obtained by substituting the innovation term in Eq.~\eqref{eq22:stateequinnovform} for ${\mathbf{e}(t)} = \mathbf{y}(t) - \mathbf{C} \mathbf{x}(t) - \mathbf{D} \mathbf{u}(t)$. By defining $q$ as the forward shift operator, \emph{i.e.},
\begin{equation}
q \mathbf{r}(t) = \mathbf{r}(t+1)
\end{equation}
for a signal $\mathbf{r}(t) \in \mathbb{R}^{n_r}$, the state equation~\eqref{eq22:stateequpredform} can be written as follows
\begin{equation}
(q \mathbf{I}_{n_x \times n_x} - \tilde{\mathbf{A}}) \mathbf{x}(t) = \tilde{\mathbf{B}} \mathbf{u}(t) + \bm{\mathcal{K}} \mathbf{y}(t).
\end{equation}
Then, by assuming that the gain matrix $\bm{\mathcal{K}}$ is built so that Proposition~\ref{prop:miniphase} is satisfied (see Theorem~\ref{th:ltikalmanfilter}), we get
\begin{equation}
\mathbf{x}(t) = (q \mathbf{I}_{n_x \times n_x} - \tilde{\mathbf{A}})^{-1} \tilde{\mathbf{B}} \mathbf{u}(t) + (q \mathbf{I}_{n_x \times n_x} - \tilde{\mathbf{A}})^{-1} \bm{\mathcal{K}} \mathbf{y}(t)
\end{equation}
or, in a different way,
\begin{equation}\label{eq22:stateinputoutputinv}
\mathbf{x}(t) = \sum_{j=1}^\infty \tilde{\mathbf{A}}^{j-1} \tilde{\mathbf{B}} \mathbf{u}(t-j) + \sum_{j=1}^\infty \tilde{\mathbf{A}}^{j-1} \bm{\mathcal{K}} \mathbf{y}(t-j)
\end{equation}
by using the Neumann series \cite{GL96}
\begin{equation}
(q \mathbf{I}_{n_x \times n_x} - \mathbf{A})^{-1} = \sum_{j=1}^\infty \mathbf{A}^{j-1} q^{-j} 
\end{equation}
where $q^{-1}$ stands for the backward shift operator. Thus, the state sequence can be constructed from past inputs and outputs, \emph{i.e.}, two full rank matrices $\bm{\Omega}_{\infty}(\tilde{\mathbf{A}},\bm{\mathcal{K}}) \in \mathbf{R}^{n_x \times \infty}$ and $\bm{\Omega}_{\infty}(\tilde{\mathbf{A}},\tilde{\mathbf{B}}) \in \mathbf{R}^{n_x \times \infty}$ exist such that
\begin{equation}
\mathbf{x}(t) = \bm{\Omega}_{\infty}(\tilde{\mathbf{A}},\bm{\mathcal{K}}) \mathbf{y}^-(t) + \bm{\Omega}_{\infty}(\tilde{\mathbf{A}},\tilde{\mathbf{B}}) \mathbf{u}^-(t) 
\end{equation}
where
\begin{subequations}
\begin{align}
 \bm{\Omega}_{\infty}(\tilde{\mathbf{A}},\bm{\mathcal{K}}) &=
\begin{bmatrix}
\cdots & \tilde{\mathbf{A}}^2 \bm{\mathcal{K}} & \tilde{\mathbf{A}} \bm{\mathcal{K}} & \bm{\mathcal{K}}  
  \end{bmatrix} \\
\bm{\Omega}_{\infty}(\tilde{\mathbf{A}},\tilde{\mathbf{B}}) &=
   \begin{bmatrix}
\cdots & \tilde{\mathbf{A}}^2 \tilde{\mathbf{B}} & \tilde{\mathbf{A}} \tilde{\mathbf{B}} & \tilde{\mathbf{B}}
  \end{bmatrix}.
\end{align}
\end{subequations}

\subsubsection{Approximated state sequence construction from finite time series}\label{para22:finitestateseqapprox}

In practice, the user usually does not have access to infinite time series. Therefore, when finite input and output sequences are handled, approximations must be used. In order to introduce these approximations, a reformulation of the model~\eqref{eq22:stateinputoutputinv} as follows is necessary
\begin{multline}\label{eq22:stateinputoutputinvp}
  \mathbf{x}(t) = \sum_{j=1}^p \tilde{\mathbf{A}}^{j-1} \tilde{\mathbf{B}} \mathbf{u}(t-j) + \sum_{j=1}^p \tilde{\mathbf{A}}^{j-1} \bm{\mathcal{K}} \mathbf{y}(t-j) \\ + \sum_{j=p + 1}^\infty \tilde{\mathbf{A}}^{j-1} \tilde{\mathbf{B}} \mathbf{u}(t-j) + \sum_{j=p + 1}^\infty \tilde{\mathbf{A}}^{j-1} \bm{\mathcal{K}} \mathbf{y}(t-j)
\end{multline}
where $p \in \mathbb{N}_*^+$ is a user-defined parameter corresponding to the available ``past''\footnote{For many subspace-based identification algorithms \cite{OM96,Kat05,VV07}, the present instant is denoted by $t$. Then, the past data is related to a user-defined integer $p$ corresponding to the time instants $[ t-p, \cdots, t-1 ]$. Similarly, a user-defined integer $f$ can be introduced in order to define the future data, \emph{i.e.}, the time instants $[ t, \cdots, t + f -1 ]$.} input and output data. By using the compact notations introduced beforehand, % Introducing the following vectors and matrices
% \begin{align}\label{eq22:stackedpastvector}
%   \mathbf{u}_{p}^-(t) &=
% \begin{bmatrix}
% \mathbf{u}(t-1) \\
% \vdots \\
% \mathbf{u}(t-p)
% \end{bmatrix}  \in \mathbb{R}^{n_u p} &   
% \mathbf{y}_{p}^-(t) &=
% \begin{bmatrix}
% \mathbf{y}(t-1) \\
% \vdots \\
% \mathbf{y}(t-p)
% \end{bmatrix} \in \mathbb{R}^{n_y p}
% \end{align}
% \begin{subequations}
% \begin{align}
%  \bm{\Lambda}_p(\tilde{\mathbf{A}} , \bm{\mathcal{K}}) &=
% \begin{bmatrix}
%  \bm{\mathcal{K}} & \tilde{\mathbf{A}} \bm{\mathcal{K}} & \tilde{\mathbf{A}}^2 \bm{\mathcal{K}} & \cdots & \tilde{\mathbf{A}}^{p - 1} \bm{\mathcal{K}}
%   \end{bmatrix} \in \mathbb{R}^{n_x \times n_y p} \\
% \bm{\Lambda}_p(\tilde{\mathbf{A}} , \tilde{\mathbf{B}}) &=
%    \begin{bmatrix}
%  \tilde{\mathbf{B}} & \tilde{\mathbf{A}} \tilde{\mathbf{B}} & \tilde{\mathbf{A}}^2 \tilde{\mathbf{B}} & \cdots & \tilde{\mathbf{A}}^{p - 1} \tilde{\mathbf{B}} 
%   \end{bmatrix} \in \mathbb{R}^{n_x \times n_u p} ,
% \end{align}
% \end{subequations}
Eq.~\eqref{eq22:stateinputoutputinvp} can be rewritten as follows
\begin{equation}
  \mathbf{x}(t) = \bm{\Omega}_p(\tilde{\mathbf{A}},\bm{\mathcal{K}}) \mathbf{y}_{p}^-(t) + \bm{\Omega}_p(\tilde{\mathbf{A}},\tilde{\mathbf{B}}) \mathbf{u}_{p}^-(t) + \bm{\upsilon}(t)
\end{equation}
where the error term $\bm{\upsilon}(t)$ is defined as
\begin{equation}
\bm{\upsilon}(t) = \sum_{j=p + 1}^\infty \tilde{\mathbf{A}}^{j-1} \tilde{\mathbf{B}} \mathbf{u}(t-j) + \sum_{j=p + 1}^\infty \tilde{\mathbf{A}}^{j-1} \bm{\mathcal{K}} \mathbf{y}(t-j) .
\end{equation}
Thus, given the observations $\mathbf{y}_{p}^-(t)$ and $\mathbf{u}_{p}^-(t)$, it is consistent to approximate the state sequence $\mathbf{x}$ as follows
\begin{equation}\label{eq22:stateapprox}
  \bar{\mathbf{x}}(t) = \bm{\Omega}_p(\tilde{\mathbf{A}},\bm{\mathcal{K}}) \mathbf{y}_{p}^-(t) + \bm{\Omega}_p(\tilde{\mathbf{A}},\tilde{\mathbf{B}}) \mathbf{u}_{p}^-(t) .
\end{equation}
It is now important to assess the quality of this linear estimate $\bar{\mathbf{x}}$ of the state sequence $\mathbf{x}$. First, the error term $\bm{\upsilon}(t) = \mathbf{x}(t) - \bar{\mathbf{x}}(t)$ is orthogonal to $\begin{bmatrix} {\mathbf{y}_{p}^-}^\top(t) & {\mathbf{u}_{p}^-}^\top(t) \end{bmatrix}^\top $ \cite{Bau03}. Second, from the system of equations corresponding to the predictor form~\eqref{eq22:sspredform} and standard recursions, it can be verified that
\begin{equation}\label{eq22:stateseqtildeA}
  \mathbf{x}(t) = \tilde{\mathbf{A}}^p \mathbf{x}(t-p) + \bm{\Omega}_p(\tilde{\mathbf{A}} , \bm{\mathcal{K}}) \mathbf{y}_{p}^-(t) + \bm{\Omega}_p(\tilde{\mathbf{A}} , \tilde{\mathbf{B}}) \mathbf{u}_{p}^-(t) .
\end{equation}
Therefore, again, the state is a linear function of past input and output signals plus a prior value of the state. Furthermore, we can show that \cite{Bau03}
\begin{multline}
  \left\| \bm{\upsilon}(t) \right\|_2 = \left\| \mathbf{x}(t) - \bm{\Omega}_p(\tilde{\mathbf{A}},\bm{\mathcal{K}}) \mathbf{y}_{p}^-(t) - \bm{\Omega}_p(\tilde{\mathbf{A}},\tilde{\mathbf{B}}) \mathbf{u}_{p}^-(t) \right\|_2 \\ = \| \tilde{\mathbf{A}}^p \mathbf{x}(t-p) \|_2 \leq \| \tilde{\mathbf{A}}^p \|_F \| \mathbf{x}(t-p) \|_2 
\end{multline}
where, for a vector $\mathbf{r}(t) \in \mathbb{R}^{n_r}$, $\left\| \mathbf{r} \right\|_2 = \sqrt{\mathbf{r}^\top \mathbf{r}}$ and, for a real square matrix $\mathbf{A}$, $\left\| \mathbf{A} \right\|_F = \sqrt{\tr{(\mathbf{A}^\top \mathbf{A})}}$ \cite{HJ90}. A direct consequence of this relation combined with the strict minimum phase condition is that the error term
\begin{itemize}
\item can be small \cite{Bau03} for a truncation index $p$ chosen ``sufficiently large'',
\item vanishes when $p \rightarrow \infty$.
\end{itemize}
Thus, the observer $\bar{\mathbf{x}}$ can be viewed as the optimal linear estimate of $\mathbf{x}$ (in the mean-square error sense) given $\mathbf{u}_{p}^-(t)$ and $\mathbf{y}_{p}^-(t)$ \cite{JW96,Kat05}. Notice however that the asymptotic case ($p \rightarrow \infty$) is only of theoretical interest (and is needed to guarantee the consistency of the estimates introduced hereafter (see Sub-Sections~\ref{para22:ls4sidopenloop} and \ref{para22:ls4sidclosedloop})) and generally cannot be satisfied in the system identification framework. Hence, in practice, the tuning parameter $p$ must be chosen 
\begin{itemize}
\item large enough in order to ensure a small error,
\item not too large in order to avoid a huge increase of the variance of the parameters when such a model is used for system identification.
\end{itemize}
A usual trade-off between bias and variance is thus necessary. This trade-off can be properly satisfied by minimizing specific Akaike information criteria such as those suggested, \emph{e.g.}, in \cite[Chapter 6 and 8]{Pet95} or in \cite{Kue05}. For a lightly damped predictor form~\eqref{eq22:sspredform}, $p$ must be chosen very large in order to satisfy $\tilde{\mathbf{A}}^i = \mathbf{0}$ for $i \geq p$. Such a situation can lead to numerical problems when, \emph{e.g.}, inversion of matrices with a size increasing with $p$ is required. However, it is interesting to point out that the error can be arbitrarily small (even equal to zero) when
\begin{itemize}
\item the system to identify is strictly deterministic,
\item an observer is introduced in order to create an additional freedom for the optimizer (similar to what was suggested in \cite{Jal93,Pal95} or more recently in \cite{HWV09b}).
\end{itemize}
Both cases are shortly described in the following two paragraphs.

%Paragraph
\paragraph{Deterministic system}

In this paragraph, let us assume that the noise sequences $\mathbf{v}$, $\mathbf{w}$ and by extension $\mathbf{e}$ are equal to zero, \emph{i.e.}, the system to identify is deterministic. Then, the system dynamics satisfy
\begin{subequations}
\begin{align}
\mathbf{x}(t+1) &= \mathbf{A} \mathbf{x}(t) + \mathbf{B} \mathbf{u}(t) \label{eq22:detstateequ} \\
\mathbf{y}(t) &= \mathbf{C} \mathbf{x}(t) + \mathbf{D} \mathbf{u}(t) \label{eq22:detoutputequ}  .
\end{align}
\end{subequations}
By iterating the state equation~\eqref{eq22:detstateequ}, it is straightforward that
\begin{equation}\label{eq22:detstate}
\mathbf{x}(t) = \mathbf{A}^{n_x} \mathbf{x}(t-n_x) + \bm{\Omega}_{n_x}(\mathbf{A},\mathbf{B}) \mathbf{u}_{n_x}^-(t) .
\end{equation}
% where 
% \begin{equation}
% \bm{\Omega}_{n_x} =
% \begin{bmatrix}
% \mathbf{B} & \mathbf{A} \mathbf{B} & \cdots & \mathbf{A}^{n_x-1} \mathbf{B}
% \end{bmatrix}
% \end{equation}
% and 
% \begin{equation}
% \mathbf{u}_{n_x}^-(t) =
% \begin{bmatrix}
% \mathbf{u}(t-1) \\
% \vdots \\
% \mathbf{u}(t-n_x)
% \end{bmatrix} .
% \end{equation}
By using combinations of Eq.~\eqref{eq22:detstateequ} and~\eqref{eq22:detoutputequ}, it is found that
\begin{equation}
\mathbf{y}_{n_x}^-(t) = \bm{\Gamma}_{n_x}(\mathbf{A},\mathbf{C}) \mathbf{x}(t-n_x) + \mathbf{H}_{n_x}(\mathbf{A},\mathbf{B},\mathbf{C},\mathbf{D}) \mathbf{u}_{n_x}^-(t) .
\end{equation}
% where $\mathbf{y}_{n_x}^+(t-n_x)$ and $\mathbf{u}_{n_x}^+(t-n_x)$ are composed of past data, \emph{i.e.},
% \begin{align}
%   \mathbf{y}_{n_x}^+(t-n_x) &= 
% \begin{bmatrix}
% \mathbf{y}(t-n_x) \\
% \vdots \\
% \mathbf{y}(t-1)
% \end{bmatrix} & 
% \mathbf{u}_{n_x}^+(t-n_x) &= 
% \begin{bmatrix}
% \mathbf{u}(t-n_x) \\
% \vdots \\
% \mathbf{u}(t-1)
% \end{bmatrix} .
% \end{align}
Now, by assuming that the system is observable, then 
\begin{equation}
  \rank{ \left( \bm{\Gamma}_{n_x}(\mathbf{A},\mathbf{C}) \right) } = n_x 
\end{equation}
and the observability matrix has, at least, $n_x$ linearly independent rows. By assuming\footnote{This assumption is satisfied for any observable MISO system. For a MIMO system, the selection of $n_x$ linearly independent rows among the $n_y f$ composing $\bm{\Gamma}_{n_x}(\mathbf{A},\mathbf{C})$ can be made as explained in \cite{MB11}.} that the first $n_x$ rows of $\bm{\Gamma}_{n_x}(\mathbf{A},\mathbf{C})$ are linearly independent,  $\bm{\Gamma}_{n_x}(\mathbf{A},\mathbf{C})[1:n_x,:]$ is a square full rank matrix\footnote{$\mathbf{A}[i:j,k:\ell]$ is the matrix composed of the coefficients present at the intersection of the rows $i$ to $j$ and the columns $k$ to $\ell$ of the matrix $\mathbf{A}$.}. Thus, we can write that
\begin{multline}\label{eq22:detpaststate}
\mathbf{x}(t-n_x) = \bm{\Gamma}_{n_x}^{-1}(\mathbf{A},\mathbf{C})[1:n_x,:] \mathbf{y}_{n_x}^-(t)[1:n_x,:] \\ - \bm{\Gamma}_{n_x}^{-1}(\mathbf{A},\mathbf{C})[1:n_x,:] \mathbf{H}_{n_x}(\mathbf{A},\mathbf{B},\mathbf{C},\mathbf{D})[1:n_x,:] \mathbf{u}_{n_x}^-(t) .
\end{multline}
Then, by gathering Eq.~\eqref{eq22:detstate} and~\eqref{eq22:detpaststate}, we get
\begin{multline}
\mathbf{x}(t) = \mathbf{A}^{n_x} \bm{\Gamma}_{n_x}^{-1}(\mathbf{A},\mathbf{C})[1:n_x,:] \mathbf{y}_{n_x}^-(t)[1:n_x,:] + \left( \bm{\Omega}_{n_x}(\mathbf{A},\mathbf{B}) \right. \\ \left. - \mathbf{A}^{n_x} \bm{\Gamma}_{n_x}^{-1}(\mathbf{A},\mathbf{C})[1:n_x,:] \mathbf{H}_{n_x}(\mathbf{A},\mathbf{B},\mathbf{C},\mathbf{D})[1:n_x,:] \right) \mathbf{u}_{n_x}^-(t) .
\end{multline}
This last equation shows that the state sequence can be formulated as a linear combination of finite past input time series and finite past output time series without any approximation. An equivalent result, \emph{i.e.}, without approximation, can be obtained for SISO ARX systems as proved in \cite{WJ94}.
 
%Paragraph
\paragraph{System with a user-defined observer}

In the general case, \emph{i.e.}, when MIMO stochastic processes are involved, the state sequence cannot be exactly recovered from finite data sets and approximations such as Eq.~\eqref{eq22:stateapprox}, must be handled. The main problem with the approximation~\eqref{eq22:stateapprox} is that the error term $\bm{\upsilon}(t) = \mathbf{x}(t) - \bar{\mathbf{x}}(t) = \tilde{\mathbf{A}}^p \mathbf{x}(t-p)$ indirectly depends on the Kalman gain matrix $\bm{\mathcal{K}}$ which, by construction,
\begin{itemize}
\item is time-invariant only asymptotically,
\item is related to the stochastic properties of the system via the matrices $\mathbf{Q}$, $\mathbf{R}$ and $\mathbf{S}$ defined in Assumption~\ref{ass:noise} and, so, cannot be freely chosen by the user,
\item requires to solve a Riccati equation.
\end{itemize}
These drawbacks make the assertion $\tilde{\mathbf{A}}^p = \mathbf{0}$ difficult to verify for a finite tuning parameter $p$. In order to circumvent this difficulty, a user-defined observer gain $\bm{\Lambda} \in \mathbb{R}^{n_x \times n_y}$ (which exists because the system is observable) is introduced in \cite{Jal93} so that the predictor form becomes 
\begin{subequations}
\begin{align}
\mathbf{x}(t+1) &= \tilde{\mathbf{A}} \mathbf{x}(t) + \tilde{\mathbf{B}} \mathbf{u}(t) + \bm{\mathcal{K}} \mathbf{y}(t) + \bm{\Lambda} (\mathbf{y}(t) - \mathbf{C} \mathbf{x}(t) - \mathbf{D} \mathbf{u}(t)) \\
\mathbf{y}(t) &= \mathbf{C} \mathbf{x}(t) + \mathbf{D} \mathbf{u}(t) + {\mathbf{e}(t)} 
\end{align}
\end{subequations}
or, more compactly,
\begin{subequations}\label{eq22:sspredformobs}
\begin{align}
\begin{split}
\mathbf{x}(t+1) &= \underbrace{(\mathbf{A} - \bm{\Lambda} \mathbf{C})}_{\breve{\mathbf{A}}} \mathbf{x}(t) + \underbrace{(\mathbf{B} - \bm{\Lambda} \mathbf{D})}_{\breve{\mathbf{B}}} \mathbf{u}(t) + \bm{\mathcal{K}} \mathbf{e}(t) + \bm{\Lambda} \mathbf{y}(t) 
\end{split} \\
\mathbf{y}(t) &= \mathbf{C} \mathbf{x}(t) + \mathbf{D} \mathbf{u}(t) + {\mathbf{e}(t)} .
\end{align}
\end{subequations}
Then, it can be shown that this model satisfies
\begin{multline}\label{eq22:stateseqbreveA}
 \mathbf{x}(t) =  \breve{\mathbf{A}}^p  \mathbf{x}(t-p) +
\bm{\Omega}_p(\breve{\mathbf{A}},\bm{\Lambda}) \mathbf{y}_p^-(t) \\ +
\bm{\Omega}_p(\breve{\mathbf{A}},\breve{\mathbf{B}}) \mathbf{u}_p^-(t) + \bm{\Omega}_p(\breve{\mathbf{A}},\bm{\mathcal{K}}) \mathbf{e}_p^-(t)
\end{multline}
where $\mathbf{e}_p^-(t)$ is defined like, \emph{e.g.}, $\mathbf{u}_p^-(t)$.
% and where the following generic notation is used
% \begin{equation}
% \bm{\Lambda}_p(\mathbf{A} , \mathbf{B}) = 
% \begin{bmatrix} 
% \mathbf{B} & \mathbf{A} \mathbf{B} & \cdots & \mathbf{A}^{p-1} \mathbf{B} \end{bmatrix}
% \end{equation}
% for two matrices $\mathbf{A}$ and $\mathbf{B}$ with adequate dimensions.
By comparing the state sequence in Eq.~\eqref{eq22:stateseqbreveA} with Eq.~\eqref{eq22:stateseqtildeA}, it is interesting to point out that
\begin{itemize}
\item the contribution of $\mathbf{x}(t-p)$ can be made arbitrarily small with the observer-predictor form~\eqref{eq22:sspredformobs} by placing the eigenvalues of the matrix $\breve{\mathbf{A}}$ in any desired configuration. This can be done with a smaller value of the user-defined parameter $p$ than when the standard predictor form~\eqref{eq22:sspredform} is used. Indeed, because the system is observable, the matrix gain $\bm{\Lambda}$ can be chosen (by the user) so that the eigenvalues of $\breve{\mathbf{A}}$ are assigned arbitrarily, \emph{e.g.}, at the origin (leading to a deadbeat observer \cite{VW99}),
\item the state sequence~\eqref{eq22:stateseqbreveA} involves past input and output signals (via the vectors $\mathbf{u}_p^-(t)$ and $\mathbf{y}_p^-(t)$) as well as past innovation terms $\mathbf{e}_p^-(t)$. As shown hereafter (see Sub-Section~\ref{para22:ls4sidopenloop}), this last term can make the estimation phase more complicated than with the standard predictor form~\eqref{eq22:sspredform}. Standard optimization methods, such as the extended least-squares algorithm \cite{SS89,Lju99}, can be adapted in order to circumvent this difficulty (see \cite{HWV09b} for details).
\end{itemize}

\begin{remark}
When the user-defined integer $p$ is chosen such that $\breve{\mathbf{A}}^i = \mathbf{0}$ (or  $\tilde{\mathbf{A}}^i = \mathbf{0}$) for $i \geq p$, the (non-zero) initial conditions can be neglected. It is then possible to focus on the steady-state solutions.
\end{remark}

% SUBSUBSECTION %
\subsubsection{To sum up}

The three previous paragraphs have introduced different ways to express the state sequence of the LTI state-space system~\eqref{eq22:ssinnovform}. By having access to this state sequence, the second step of the linear regression-based subspace identification methods addresses the estimation of the state-space matrices $\left(\mathbf{A}, \mathbf{B}, \mathbf{C}, \mathbf{D} \right)$ and possibly the Kalman gain $\bm{\mathcal{K}}$ (up to a similarity transformation) given realizations $\left\{\mathbf{u}(t)\right\}^N_{t=1}$ and $\left\{\mathbf{y}(t)\right\}^N_{t=1}$ of the input and output signals on a finite but sufficiently wide time horizon $N$. This problem is investigated in Sub-Section~\ref{para22:ls4sidopenloop} and \ref{para22:ls4sidclosedloop}. More precisely, solutions dedicated to open-loop systems are first described\footnote{This overview can be considered as a combination of results available mainly in \cite{Pet95,Bau03,Bau05}. Notice that {D. Bauer} also surveyed the most relevant asymptotic properties of the subspace estimators in \cite{Bau03,Bau05}.} in Sub-Section~\ref{para22:ls4sidopenloop}. Extensions to systems operating in closed-loop are then introduced in Sub-Section~\ref{para22:ls4sidclosedloop}.

%%%%%%%%%%%%%%%%%%%%%%%%%%% SUBSECTION %%%%%%%%%%%%%%%%%%%%%%%%%%
\subsection[Subspace ident. using linear regression for open-loop systems]{Subspace-based identification using linear regression for systems operating in open-loop}\label{para22:ls4sidopenloop}

% Let us define the generic stacked vector
% \begin{equation}
% \mathbf{r}_{f}^+(t) =
% \begin{bmatrix}
% \mathbf{r}(t) \\
% \vdots \\
% \mathbf{r}(t + f - 1)
% \end{bmatrix} \in \mathbb{R}^{n_r f}
% \end{equation}
% for any vector $\mathbf{r}(t) \in \mathbb{R}^{n_r}$ where $f \in \mathbb{N}^+_*$ is a user-defined integer\footnote{Most of the equations introduced hereafter are accurate for any value of $f$. However, for rank constraint reasons, it is assumed that $f$ is chosen such that $f \geq n_x$.}. This way of sorting data is the starting point of most of the subspace-based identification methods available in the literature. More precisely, by iterating the equations composing Eq.~\eqref{eq22:ssinnovform}, it is straightforward to obtain the following data equation
For $f \in \mathbb{N}^+_*$ a user-defined integer\footnote{Most of the equations introduced hereafter are accurate for any value of $f$. However, for rank constraint reasons, it is assumed that $f$ is chosen such that $f \geq n_x$.}, the starting point of most of the subspace-based identification methods available in the literature is the following data equation \cite{Vib95}
\begin{multline}\label{eq22:vectordataequ}
\mathbf{y}_{f}^+(t) = \bm{\Gamma}_{f}(\mathbf{A}, \mathbf{C}) \mathbf{x}(t) + \mathbf{H}_{f}(\mathbf{A}, \mathbf{B}, \mathbf{C}, \mathbf{D}) \mathbf{u}_{f}^+(t)  \\ + \mathbf{H}_{f}(\mathbf{A}, \bm{\mathcal{K}}, \mathbf{C}, \mathbf{I}_{n_y}) \mathbf{e}_{f}^+(t) .
\end{multline}
Because this sub-section is dedicated to the identification of systems working in open-loop, the following assumptions will be satisfied hereafter.
\begin{assumption}\label{ass:openstability}
The system represented by the LTI state-space form~\eqref{eq22:ssinnovform} is (asymptotically) stable, \emph{i.e.},
  \begin{equation}
    \lambda_{\textup{max}} (\mathbf{A}) < 1 .
  \end{equation}
\end{assumption}
\begin{assumption}\label{ass:miniphase}
The system represented by the LTI state-space form~\eqref{eq22:ssinnovform} is strict minimum phase, \emph{i.e.},
  \begin{equation}
    \lambda_{\textup{max}} (\mathbf{A} - \bm{\mathcal{K}} \mathbf{C}) < 1 .
  \end{equation}
\end{assumption}
\begin{assumption}\label{ass:noiseinputindep}
  The noise terms $\mathbf{v}$ and $\mathbf{w}$ are zero-mean and statistically independent of the input $\mathbf{u}$. By extension, the innovation $\mathbf{e}$ is zero-mean and statistically independent of the input $\mathbf{u}$.
\end{assumption}

% SUBSUBSECTION %
\subsubsection{Unconstrained least-squares solutions}\label{para22:unconstlssol}

The central idea of the regression-based techniques is to substitute the approximation~\eqref{eq22:stateseqtildeA} (or~\eqref{eq22:stateseqbreveA}) for the state sequence $\mathbf{x}$ in Eq.~\eqref{eq22:vectordataequ}. More precisely, after straightforward manipulations of the matrices involved in Eq.~\eqref{eq22:stateseqtildeA} (or Eq.~\eqref{eq22:stateseqbreveA}) and Eq.~\eqref{eq22:vectordataequ}, the combination of Eq.~\eqref{eq22:stateseqtildeA} (or Eq.~\eqref{eq22:stateseqbreveA}) and Eq.~\eqref{eq22:vectordataequ} leads to the following compact equation \cite{PSD96,JW98,Qin06}
\begin{equation}\label{eq22:linregcompvectordataequ}
\mathbf{y}_{f}^+(t) = \underbrace{\bm{\Gamma}_{f}(\mathbf{A}, \mathbf{C}) \bm{\Omega}_p(\bm{\mathcal{A}},\bm{\mathcal{B}})}_{\mathbf{L}_{f,p}} \mathbf{z}_{p}^-(t) + \underbrace{\mathbf{H}_{f}(\mathbf{A},\mathbf{B},\mathbf{C},\mathbf{D})}_{\mathbf{H}_{f}^{ol,u}} \mathbf{u}_{f}^+(t) + \mathbf{n}_{f}^+(t)
\end{equation}
where 
\begin{equation}
  \mathbf{z}(t) =
  \begin{bmatrix}
    \mathbf{y}(t) \\
\mathbf{u}(t)
  \end{bmatrix} \in \mathbb{R}^{n_y + n_u}
\end{equation}
and, by extension,  $\mathbf{z}_{p}^- \in \mathbb{R}^{(n_u + n_y) p} $ and where $\bm{\mathcal{A}}$, $\bm{\mathcal{B}}$ and $\mathbf{n}_{f}^+(t)$ are defined as follows
\begin{align}
\bm{\mathcal{A}} &= \tilde{\mathbf{A}} & \bm{\mathcal{B}} = 
\begin{bmatrix}
\bm{\mathcal{K}} & \tilde{\mathbf{B}}
\end{bmatrix}
\end{align}
\begin{equation}\label{eq22:tildeAnoise}
  \mathbf{n}_{f}^+(t) =  \underbrace{\mathbf{H}_{f}(\mathbf{A}, \bm{\mathcal{K}}, \mathbf{C}, \mathbf{I}_{n_y})}_{\mathbf{H}_{f}^{ol,e}} \mathbf{e}_{f}^+(t) + \bm{\Gamma}_{f}(\mathbf{A}, \mathbf{C}) \tilde{\mathbf{A}}^p \mathbf{x}(t-p)
\end{equation}
when Eq.~\eqref{eq22:stateseqtildeA} is used or
\begin{align}
\bm{\mathcal{A}} &= \breve{\mathbf{A}} & \bm{\mathcal{B}} = 
\begin{bmatrix}
\bm{\Lambda} & \breve{\mathbf{B}}
\end{bmatrix}
\end{align}
\begin{equation}\label{eq22:breveAnoise}
  \mathbf{n}_{f}^+(t) =  \mathbf{H}_{f}^{ol,e} \mathbf{e}_{f}^+(t) + \bm{\Gamma}_{f}(\mathbf{A}, \mathbf{C}) \breve{\mathbf{A}}^p \mathbf{x}(t-p) + \bm{\Gamma}_{f}(\mathbf{A}, \mathbf{C})
\bm{\Omega}_p(\breve{\mathbf{A}},\bm{\mathcal{K}}) \mathbf{e}_p^-(t)
\end{equation}
when  Eq.~\eqref{eq22:stateseqbreveA} is favored. Furthermore, by assuming that
\begin{itemize}
\item the assumptions~\ref{ass:minimal}, \ref{ass:miniphase} and \ref{ass:noiseinputindep} are satisfied,
\item the user-defined integer $p$ and $f$ are chosen at least greater than or equal to $n_x$ and sufficiently large to ensure that the following assumptions are satisfied
\begin{assumption}\label{ass:tildeAnilpotent}
The matrix $\tilde{\mathbf{A}}$ is nilpotent of degree $p$, \emph{i.e.}, $\tilde{\mathbf{A}}^i = \mathbf{0}$ for $i \geq p$,
\end{assumption}
\begin{assumption}\label{ass:breveAnilpotent}
The matrix $\breve{\mathbf{A}}$ is nilpotent of degree $p$, \emph{i.e.}, $\breve{\mathbf{A}}^i = \mathbf{0}$ for $i \geq p$,
\end{assumption}
 \end{itemize}
then
\begin{itemize}
\item the terms $\tilde{\mathbf{A}}^p \mathbf{x}(t-p)$ and $\breve{\mathbf{A}}^p  \mathbf{x}(t-p)$ are negligible,
\item the noise term $ \mathbf{H}_{f}^{ol,e} \mathbf{e}_{f}^+(t)$ is uncorrelated with $\mathbf{z}_{p}^-(t)$ as well as $\mathbf{u}_{f}^-(t)$,
\item the following geometric properties are satisfied \cite{JW98,Bau05}
 \begin{subequations}
\begin{align}
\rank{\left( \mathbf{L}_{f,p} \right)} &= n_x \\
\col{\left(  \mathbf{L}_{f,p} \right)} &= \col{\left( \bm{\Gamma}_{f}(\mathbf{A},\mathbf{C}) \right)}
\end{align}
\end{subequations}
where $\col{\left( \bullet \right)}$ is the space generated by the columns of $\bullet$.
\end{itemize}
These latter structural properties could be taken into account for the subspace estimation step. 

Before describing several subspace-based identification algorithms developed to estimate the subspaces $\mathbf{L}_{f,p}$ and $\mathbf{H}_{f}^{ol,e}$ (under constraints), the data equations introduced must be slightly modified. More precisely, given input and output data on a finite but sufficiently wide time horizon, future and past Hankel matrices of the signals involved in Eq.~\eqref{eq22:linregcompvectordataequ} can be defined\footnote{$M$ is defined in a way to be compatible with the full number of input-output measurements.} (see Eq.~\eqref{eq22:futhankelmat} for an explicit definition) and the vector data equation~\eqref{eq22:linregcompvectordataequ} becomes the matrix data equation\footnote{The time index is dropped when this index is not necessary for the understanding of the equations.}
\begin{equation}\label{eq22:linregcompmatdataequ}
\mathbf{Y}_{f,M}^+ = \mathbf{L}_{f,p} \mathbf{Z}_{p,M}^- + \mathbf{H}_{f}^{ol,u} \mathbf{U}_{f,M}^+ +  \mathbf{N}_{f,M}^+
\end{equation}
with $\mathbf{Z}_{p,M}^- \in \mathbb{R}^{(n_u + n_y) f \times M}$.
This matrix data equation lies at the core of the subspace-based identification idea and the dedicated algorithms \cite{Bau05}.

\begin{remark}
  In this paper, it has been chosen to work with Hankel matrices. This choice is shared by many standard subspace-based identification methods \cite{Vib95,OM96,VV07}. Recently, rather than working with (infinite) Hankel matrices, a stochastic framework based on specific Hilbert spaces of random variables generated by the signals involved in the identification problem has been considered by {D. Bauer} \cite{Bau03,Bau05}, {A. Chiuso} and {G. Picci} \cite{CP02,CP03} and {T. Katayama} \cite{Kat05} (to cite only the main contributions). By doing so, a new geometric interpretation of the subspace-based identification approach can be suggested (see, \emph{e.g.}, \cite{CP05,Kat05} for important details concerning the definitions and the notations involved in this stochastic framework). As claimed in \cite{CP02}, ``the stochastic realization theory lies at the ground of the subspace-based identification''. Such a stochastic setting has indeed several advantages. First, it can be really beneficial for the derivation of some asymptotic properties of the subspace estimators \cite{Bau03}. Indeed, as shown by {A. Chiuso} in most of his contributions \cite{CP04,CP04c,Chi07b,Chi10} dedicated to the subspace-based identification, a generalization of the well-established theory for time-series can be performed in the subspace-based identification framework (with exogenous inputs) thanks to the close link between the stochastic realization theory and several geometric tools involved in subspace-based identification. Most of the steps used in the standard subspace-based identification methods as well as some linear algebra tools (\emph{e.g.}, the oblique projection) can be viewed as sampled versions of certain operations used in the stochastic realization theory \cite{CP02}. Second, from a practical point of view, the use of vector equations (instead of matrix equations) can be convenient when, \emph{e.g.}, huge data sets are involved or missing values as well as outliers are present in the data sets. Some of these benefits have been highlighted in \cite{Bau98,Bau03}. These interesting theoretical and practical properties can be translated into the sampled data matrix framework considered hereafter by noticing that a Hilbert space related to a zero-mean random variable $\mathbf{r}$ (used in the stochastic realization theory) can be replaced by the row space of $\mathbf{R}_{p,M}^-$ and $\mathbf{R}_{f,M}^+$ (see \cite{Kat05} for details).
\end{remark}

As claimed previously and shown, \emph{e.g.}, in \cite{HWV09b}, the estimation of the matrices $\mathbf{L}_{f,p}$ and $\mathbf{H}_{f}^{ol,u}$ from Eq.~\eqref{eq22:linregcompmatdataequ} is more or less complex according to the state sequence approximation used in the regression problem. Indeed, according to the use of Eq.~\eqref{eq22:stateseqtildeA} or Eq.~\eqref{eq22:stateseqbreveA} in Eq.~\eqref{eq22:vectordataequ}, two different noise terms can be deduced (see Eq.~\eqref{eq22:tildeAnoise} and Eq.~\eqref{eq22:breveAnoise}). Now, when Assumption~\ref{ass:tildeAnilpotent} and Assumption~\ref{ass:breveAnilpotent} are satisfied, the noise terms~\eqref{eq22:tildeAnoise} and~\eqref{eq22:breveAnoise} differ from the presence of a past innovation stacked vector $\mathbf{e}_p^-$. The problem with this noise contribution is that $\mathbf{e}_p^-$ can be correlated with the past output data involved in  $\mathbf{z}_p^-$. Thus, a standard least-squares optimization applied to Eq.~\eqref{eq22:linregcompmatdataequ} with $\mathbf{N}_{f,M}^+$ built from Eq.~\eqref{eq22:breveAnoise} will lead to biased estimates. In order to circumvent this difficulty, the application of the well-known extended least-squares technique \cite{Lju99} or a dedicated instrumental variable method \cite{SS89} can be considered. For instance, a recursive extended least-squares algorithm is suggested in \cite{HWV09b} for systems operating in closed-loop. Because the aforementioned estimation schemes can be viewed as well-established extensions of the ordinary least-squares, the rest of this section will only focus on the linear problem~\eqref{eq22:linregcompmatdataequ} involving
\begin{equation}\label{eq22:tildeAnoise2}
  \mathbf{n}_{f}^+(t) =  \mathbf{H}_{f}(\mathbf{A}, \bm{\mathcal{K}}, \mathbf{C}, \mathbf{I}_{n_y}) \mathbf{e}_{f}^+(t) 
\end{equation}
 as the residuals. The reader interested by the extended least-squares approach is advised to study \cite{HWV09b} and the references therein. 

By recalling that this vector $\mathbf{n}_{f}^+$ (and by extension the residual matrix $\mathbf{N}_{f,M}^+$) is uncorrelated with $\mathbf{z}_{p}^-$ and $\mathbf{u}_{f}^+$, a natural way to estimate the matrices $\mathbf{L}_{f,p}$ and $\mathbf{H}_{f}^{ol,u}$ consists in minimizing the least-squares cost function
\begin{equation}\label{eq22:lscriterionol}
V(\mathbf{L}_{f,p},\mathbf{H}_{f}^{ol,u}) = 
\left\| \mathbf{Y}_{f,M}^+ - 
\begin{bmatrix}
\mathbf{L}_{f,p} & \mathbf{H}_{f}^{ol,u}
\end{bmatrix}
\begin{bmatrix}
\mathbf{Z}_{p,M}^- \\
\mathbf{U}_{f,M}^+
\end{bmatrix} \right\|_F^2 .
\end{equation}
This criterion measures the quality of the prediction of the future outputs $\mathbf{Y}_{f,M}^+$ from a linear combination of the past input and output data $\mathbf{Z}_{p,M}^-$ and the future inputs $\mathbf{U}_{f,M}^+$ \cite{OM96}. Many solutions (and several alternative versions) are available in the literature \cite{PSD96,JW96,QL03} in order to solve such a least-squares regression. The main steps composing these contributions are summed up in the following. The interested reader can study \cite{Bau05,Qin06} and the references cited in these articles in order to complete this brief overview.

\begin{remark}
In the following, several (constrained) least-squares estimates are introduced. Only theoretical solutions for these least-squares optimization problems are given. The efficient implementation of these solutions is not addressed in this paper. A reliable implementation of a least-squares algorithm requires the use of robust numerical tools such as, \emph{e.g.}, the RQ factorization. The interested reader can study, \emph{e.g.}, \cite[Chapter 4]{SS89}, \cite{GL96} or \cite[Chapter 2]{VV07} for details about the computational aspects.
\end{remark}

%Paragraph
\paragraph{Ordinary least-squares solutions}\label{para22:ordlssol}

A standard way to minimize the cost function~\eqref{eq22:lscriterionol} consists in using an ordinary least-squares estimate \cite{PSD96}
\begin{equation}\label{eq22:olssolol1}
\begin{bmatrix}
\hat{\mathbf{L}}_{f,p} & \hat{\mathbf{H}}_{f}^{ol,u}
\end{bmatrix} =
\mathbf{Y}_{f,M}^+ 
\begin{bmatrix}
\mathbf{Z}_{p,M}^- \\
\mathbf{U}_{f,M}^+
\end{bmatrix}^{\top}
\left(
 \begin{bmatrix}
\mathbf{Z}_{p,M}^- \\
\mathbf{U}_{f,M}^+
\end{bmatrix}
\begin{bmatrix}
\mathbf{Z}_{p,M}^- \\
\mathbf{U}_{f,M}^+
\end{bmatrix}^{\top}
\right)^{-1} 
\end{equation}
because, under open-loop conditions,
\begin{equation}
\lim_{M \rightarrow \infty} \frac{1}{M} \mathbf{N}_{f,M}^+ \begin{bmatrix} {\mathbf{U}_{f,M}^+}^\top & {\mathbf{Z}_{p,M}^-}^\top \end{bmatrix} = \mathbf{0} .
\end{equation}
This solution requires that
\begin{equation}
  \rank{\left( \lim_{M \rightarrow \infty}  \begin{bmatrix}
\mathbf{Z}_{p,M}^- \\
\mathbf{U}_{f,M}^+
\end{bmatrix}
\begin{bmatrix}
\mathbf{Z}_{p,M}^- \\
\mathbf{U}_{f,M}^+
\end{bmatrix}^{\top} \right)} = n_u (p+f) + n_y p .
\end{equation}
This rank constraint can be satisfied under mild conditions on the input signals $\mathbf{u}$ acting on the system (see \cite[Lemma 9.9]{VV07} for details). Alternatively, because the knowledge of $\hat{\mathbf{H}}_{f}^{ol,u}$ can be useless for the extraction of the state-space matrices estimates, this least-squares problem can be broken down into two steps \cite{PSD96,JW96}:
\begin{itemize}
\item elimination of the forced response by applying an orthogonal projection of the row space of $\mathbf{Y}_{f,M}^+$ onto the complement of the row space of $\mathbf{U}_{f,M}^+$, \emph{i.e.},
\begin{equation}
\mathbf{Y}_{f,M}^+ \mathbf{\Pi}^{\bot}_{\mathbf{U}_{f,M}^+} = \mathbf{L}_{f,p} \mathbf{Z}_{p,M}^- \mathbf{\Pi}^{\bot}_{\mathbf{U}_{f,M}^+} + \mathbf{N}_{f,M}^+ \mathbf{\Pi}^{\bot}_{\mathbf{U}_{f,M}^+}
\end{equation}
where\footnote{This projection can be computed in a stable and efficient way by using a RQ factorization \cite{VV07}.}
\begin{equation}\label{eq22:orthproj}
\mathbf{\Pi}^{\bot}_{\mathbf{U}_{f,M}^+} = \mathbf{I}_{M \times M} - {\mathbf{U}_{f,M}^+}^\top \left( \mathbf{U}_{f,M}^+ {\mathbf{U}_{f,M}^+}^\top \right)^{-1} \mathbf{U}_{f,M}^+ .
\end{equation}
This projection requires that the matrix $\mathbf{U}_{f,M}^+$ has full row rank \cite{VV07}, 
\item estimation of the subspace $\hat{\mathbf{L}}_{f,p}$ via a least-squares solution
\begin{equation}\label{eq22:olssolol2}
\hat{\mathbf{L}}_{f,p} = \mathbf{Y}_{f,M}^+ \mathbf{\Pi}^{\bot}_{\mathbf{U}_{f,M}^+} {\mathbf{Z}_{p,M}^-}^{\top} \left( \mathbf{Z}_{p,M}^- \mathbf{\Pi}^{\bot}_{\mathbf{U}_{f,M}^+} {\mathbf{Z}_{p,M}^-}^\top \right)^{-1} .
\end{equation}
\end{itemize}
By straightforward calculations, it can be shown that both solutions~\eqref{eq22:olssolol1} and~\eqref{eq22:olssolol2} lead to the same estimate of $\mathbf{L}_{f,p}$.

As usual in the subspace-based identification framework, several authors have suggested other solutions to the least-squares problem~\eqref{eq22:lscriterionol} by introducing different weighting matrices pre-multiplying and/or post-multiplying the solutions described beforehand. {D. Bauer} addresses this problem in \cite{Bau05} and undertakes a comparison of the main techniques developed during the 1990's.

By recalling that the noise term $\mathbf{N}_{f,M}^+$ is equal to $ \mathbf{H}_{f}^{ol,e} \mathbf{E}_{f,M}^+$, the residuals are not white. Thus, the least-squares estimators introduced previously are not minimum variance unbiased estimators. If an optimal estimator is sought, the standard solution for this correlation problem consists in resorting to a global least-squares method. This technique requires a prior estimate of the covariance matrix of the residuals and is, by construction, iterative. According to the studies available in \cite{Pet95,PSD96}, such an iterative solution can slightly improve the quality of the estimates of the state-space matrices computed from $\hat{\mathbf{L}}_{f,p}$. However, to the author's point of view, because the estimation of $\mathbf{L}_{f,p}$ is only an intermediate step, the use of a more complicated algorithm (than a standard least-squares one) should be restricted to the regression problems handling the observer-predictor form~\eqref{eq22:sspredformobs} and the matrix $\breve{\mathbf{A}}$. Notice indeed that the residuals considered up until now are theoretically equal to $ \mathbf{H}_{f}^{ol,e} \mathbf{E}_{f,M}^+ + \bm{\Gamma}_{f}(\mathbf{A},\mathbf{C}) \tilde{\mathbf{A}}^p \begin{bmatrix} \mathbf{x}(t-p) & \cdots & \mathbf{x}(t+M-p-1) \end{bmatrix}$. As shown in \cite{HWV09b}, verifying Assumption~\ref{ass:tildeAnilpotent} in order to ensure an unbiased estimation requires to choose a large value\footnote{With the simulation example considered in \cite{HWV09b}, the past window is at least three times larger with $\tilde{\mathbf{A}}$ than with $\breve{\mathbf{A}}$.} for the integer $p$. Now, it is well-known that increasing the value of $p$ leads to an increase of the estimate variance. Thus, the development of an algorithm able to yield a minimum variance unbiased estimator for an estimation problem characterized by estimates with a large variance seems to be not so essential. This variance problem should be less annoying with the observer-predictor form~\eqref{eq22:sspredformobs} because the user controls the dynamic of $\breve{\mathbf{A}}$.

\begin{remark}
At this stage, it is interesting to emphasize the link between this approach and the standard subspace-based identification algorithms such as the MOESP class of methods \cite{VD92,VD92b,Ver93,Ver94,CV97} or the N4SID algorithms \cite{OM96}. 

First, let us consider the PO-MOESP algorithm \cite{Ver94}. For this algorithm, a dedicated RQ factorization is introduced in order to apply an orthogonal projection as well as an instrumental variable. More precisely, the following RQ
\begin{equation}
\begin{bmatrix}
\mathbf{U}_{f,M}^+ \\
\mathbf{Z}_{p,M}^- \\
\mathbf{Y}_{f,M}^+
\end{bmatrix} =
\begin{bmatrix}
\mathbf{R}_{11} & \mathbf{0} & \mathbf{0} \\
\mathbf{R}_{21} & \mathbf{R}_{22} & \mathbf{0} \\
\mathbf{R}_{31} & \mathbf{R}_{32} & \mathbf{R}_{33}
\end{bmatrix}
\begin{bmatrix}
\mathbf{Q}_{1} \\
\mathbf{Q}_{2} \\
\mathbf{Q}_{3}
\end{bmatrix}
\end{equation}
is considered where  ${\mathbf{R}}_{11} \in \mathbb{R}^{n_{u} f \times n_{u} f}$, ${\mathbf{R}}_{21} \in \mathbb{R}^{(n_u + n_y) p \times n_{u} f}$, ${\mathbf{R}}_{31} \in \mathbb{R}^{n_{y} f \times n_{u} f}$, ${\mathbf{R}}_{22} \in \mathbb{R}^{(n_u + n_y) p \times (n_u + n_y) p}$, ${\mathbf{R}}_{32} \in \mathbb{R}^{n_{y} f \times (n_u + n_y) p}$ and ${\mathbf{R}}_{33} \in \mathbb{R}^{n_{y} f \times n_{y} f}$ are lower triangular matrices and  ${\mathbf{Q}}_{1} \in \mathbb{R}^{n_{u} f \times M}$, ${\mathbf{Q}}_{2} \in \mathbb{R}^{k p \times M}$ and ${\mathbf{Q}}_{3} \in \mathbb{R}^{n_{y} f \times M}$ are orthogonal matrices. Then, by straightforward calculations, it can be shown that \cite{Vib95}
\begin{equation}
{\mathbf{R}}_{32} {\mathbf{Q}}_{2} = \mathbf{Y}_{f,M}^{+} \mathbf{\Pi}^{\bot}_{\mathbf{U}_{f,M}^+} {\mathbf{Z}_{p,M}^-}^{\top} \left( \mathbf{Z}_{p,M}^- \mathbf{\Pi}^{\bot}_{\mathbf{U}_{f,M}^+} {\mathbf{Z}_{p,M}^-}^{\top} \right)^{-1} \mathbf{Z}_{p,M}^- \mathbf{\Pi}^{\bot}_{\mathbf{U}_{f,M}^+} .
\end{equation}
By comparing this equation with Eq.~\eqref{eq22:olssolol2}, it is obvious that 
\begin{equation}
{\mathbf{R}}_{32} {\mathbf{Q}}_{2} = \hat{\mathbf{L}}_{f,p} \bm{\mathcal{W}}_r
\end{equation}
with the weighting matrix $\bm{\mathcal{W}}_r =  \mathbf{Z}_{p,M}^- \mathbf{\Pi}^{\bot}_{\mathbf{U}_{f,M}^+}$. This equality can be summarized algebraically as follows
\begin{equation}
\col{\left( \mathbf{R}_{32} \right)} = \col{\left( \hat{\mathbf{L}}_{f,p} \right)} .
\end{equation}
Therefore, in addition to sharing the same mathematical tools\footnote{Keep in mind that a least-squares minimization involves RQ factorizations.}, both methods are deeply correlated from a theoretical point of view.

Second, by defining the oblique projection of the row space of $\mathbf{Y}_{f,M}^+$ along the row space of $\mathbf{U}_{f,M}^+$ on the row space of $\mathbf{Z}_{p,M}^-$ as follows\footnote{${\bullet}^\dag$ stands for the Moore-Penrose pseudo-inverse of the matrix $\bullet$ \cite{GL96}.} \cite[Chapter 1]{OM96}
\begin{multline}
  \mathbf{Y}_{f,M}^+ {/_{\mathbf{U}_{f,M}^+}} \mathbf{Z}_{p,M}^- =
 \mathbf{Y}_{f,M}^+ \mathbf{\Pi}^{\bot}_{\mathbf{U}_{f,M}^+} \left( \mathbf{Z}_{p,M}^- \mathbf{\Pi}^{\bot}_{\mathbf{U}_{f,M}^+} \right)^\dag \mathbf{Z}_{p,M}^- \\
= \mathbf{Y}_{f,M}^+ \mathbf{\Pi}^{\bot}_{\mathbf{U}_{f,M}^+} {\mathbf{Z}_{p,M}^-}^{\top} \left( \mathbf{Z}_{p,M}^- \mathbf{\Pi}^{\bot}_{\mathbf{U}_{f,M}^+} {\mathbf{Z}_{p,M}^-}^\top \right)^{-1} \mathbf{Z}_{p,M}^-
\end{multline}
if $\mathbf{Z}_{p,M}^- \mathbf{\Pi}^{\bot}_{\mathbf{U}_{f,M}^+}$ has full row rank, it is obvious that this oblique projection can be directly related to the least-squares solution~\eqref{eq22:olssolol2}. Indeed, by knowing the estimate $\hat{\mathbf{L}}_{f,p}$, the optimal prediction (in the least-squares sense) of the future Hankel matrix from the past input and output data satisfies
\begin{equation}
\begin{split}
  \hat{\mathbf{Y}}_{f,M}^+ &= \hat{\mathbf{L}}_{f,p} \mathbf{Z}_{p,M}^- \\
&=  \mathbf{Y}_{f,M}^+ \mathbf{\Pi}^{\bot}_{\mathbf{U}_{f,M}^+} {\mathbf{Z}_{p,M}^-}^{\top} \left( \mathbf{Z}_{p,M}^- \mathbf{\Pi}^{\bot}_{\mathbf{U}_{f,M}^+} {\mathbf{Z}_{p,M}^-}^\top \right)^{-1} \mathbf{Z}_{p,M}^- \\
&= \mathbf{Y}_{f,M}^+ {/_{\mathbf{U}_{f,M}^+}} \mathbf{Z}_{p,M}^-
\end{split}
\end{equation}
\emph{i.e.}, is exactly equal to the oblique projection of the row space of $\mathbf{Y}_{f,M}^+$ along the row space of $\mathbf{U}_{f,M}^+$ on the row space of $\mathbf{Z}_{p,M}^-$. This observation is the main step of the combined deterministic-stochastic subspace-based identification procedure developed by {P. Van Overschee} and {B. De Moor} and known as the N4SID algorithm \cite[Chapter 4]{OM96}. This observation is also the core of the Conditional Canonical Correlation Analysis algorithm developed by {T. Katayama} and explained in \cite[Chapter 10]{Kat05} (see Eq. (10.23)) as well as the recent {A. Chiuso}'s developments and extensions to the systems operating in closed-loop (see \cite{CP03,CP05} for details concerning the importance of the oblique projection in stochastic realization theory).

\end{remark}

% SUBSUBSECTION %
\subsubsection{Constrained least-squares solutions}\label{para22:constleastsquaressol}

Although the solutions~\eqref{eq22:olssolol1} and~\eqref{eq22:olssolol2} are asymptotically reliable and accurate \cite[Chapter 7]{Pet95}, they suffer from drawbacks the user must be aware of. Most of them are listed in \cite{Bau03}. The most important ones are probably the following ones.
\begin{itemize}
\item These solutions often require to fix a quite large past window in order to satisfy Assumption~\ref{ass:tildeAnilpotent}. Thus, in practice, with a finite value of $p$, biased estimates are generally obtained.
\item They do not take into account the rank constraint $\rank{\left( \mathbf{L}_{f,p} \right)} = n_x$ as well as the Toeplitz structure of $\mathbf{H}_{f}^{ol,u}$. Indeed, the least-squares solutions~\eqref{eq22:olssolol1} and~\eqref{eq22:olssolol2} are full rank matrices.
\end{itemize}
While the bias problem\footnote{The reader must remember that using an observer-predictor form can be a solution to solve this problem because, by choosing the observer gain $\bm{\Lambda}$ correctly, Assumption~\ref{ass:breveAnilpotent} can be satisfied with a small value of $p$.} cannot be solved theoretically without imposing $p \rightarrow \infty$, the rank constraint as well as the structure of $\mathbf{H}_{f}^{ol,u}$  can be taken into account in the least-squares optimization.

\paragraph{Improvements for the calculation of  $\mathbf{H}_{f}^{ol,u}$}\label{para22:improvedcalcHf}

As far as the Toeplitz structure of $\mathbf{H}_{f}^{ol,u}$ is concerned, three main solutions are available in the literature \cite{PSD96,Shi01,QL03b}. 

The first one is based on a rewritting of the least-squares problem~\eqref{eq22:lscriterionol} by applying the vectorization operator \cite{GL96} in order to remove the zero entries in the Toeplitz matrix $\mathbf{H}_{f}^{ol,u}$. More precisely, by vectorizing Eq.~\eqref{eq22:linregcompmatdataequ} and by using standard properties of this operator, it holds that
\begin{equation}
\vect{\left( \mathbf{Y}_{f,M}^+ \right)} = 
\left(
\begin{bmatrix}
\mathbf{Z}_{p,M}^- \\
\mathbf{U}_{f,M}^+
\end{bmatrix} \otimes \mathbf{I} \right)
\vect{ \left(
\begin{bmatrix}
\mathbf{L}_{f,p}  & \mathbf{H}_{f}^{ol,u}
\end{bmatrix}
\right)}
 + \vect{\left( \mathbf{N}_{f,M}^+ \right)} .
\end{equation}
Then, a matrix $\bm{\Pi}$ (composed of zeros and ones) can be constructed such that \cite{PSD96}
\begin{multline}
\bm{\Pi} \
\vect{\left(
\begin{bmatrix}
\mathbf{D} & \mathbf{C}\mathbf{B} & \cdots & \mathbf{C}\mathbf{A}^{f-2}\mathbf{B}
\end{bmatrix}
\right)} \\
= \vect{\left(
\begin{bmatrix}
\mathbf{D} & \mathbf{0} & \cdots & \mathbf{0}\\
\mathbf{C}\mathbf{B} & \mathbf{D} & \cdots &\mathbf{0}\\
\vdots & \ddots & \ddots &\vdots\\
\mathbf{C}\mathbf{A}^{f-2}\mathbf{B} & \cdots & \mathbf{C}\mathbf{B} & \mathbf{D}
\end{bmatrix}
\right)} .
\end{multline}
The combination of both previous equations leads to a constrained least-squares estimate which can be solved by using dedicated tools (see \cite{PSD96} for details).

The second solution, also available in \cite{PSD96}, is based on a two-step procedure. By assuming that estimates of $\left( \mathbf{A}, \mathbf{B}, \mathbf{C}, \mathbf{D} \right)$ are available (for instance by using the ordinary least-squares approach introduced beforehand), estimates of the Markov parameters composing the Toeplitz matrix $\mathbf{H}_{f}^{ol,u}$ can be constructed and, by extension, an estimate of $\mathbf{H}_{f}^{ol,u}$ can be built. Then, the matrix $\mathbf{L}_{f,p}$ can be estimated by knowing $\hat{\mathbf{H}}_{f}^{ol,u}$ from the least-squares criterion
\begin{equation}
\bar{V}(\mathbf{L}_{f,p}) = 
\left\| \mathbf{Y}_{f,M}^+ - \hat{\mathbf{H}}_{f}^{ol,u} \mathbf{U}_{f,M}^+
- \mathbf{L}_{f,p} \mathbf{Z}_{p,M}^- \right\|_F^2 .
\end{equation}
In order to improve the estimation, this iterative technique can be repeated many times. As shown in the simulation examples of \cite{Pet95}, this approach is efficient and does not require many runs to yield accurate estimates.

The third technique, initially introduced in \cite{QL03b}, is a parallel reformulation of the least-squares problem. This reformulation allows the suppression of the non-causal terms which appear in the matrix data equation. In order to see this feature, let us consider the following block-wise decompositions
\begin{subequations}
\begin{align}
\mathbf{U}_{f,M}^+ &= 
\begin{bmatrix}
{^1}\mathbf{U}_{f,M}^+ \\
{^2}\mathbf{U}_{f,M}^+ \\
\vdots \\
{^f}\mathbf{U}_{f,M}^+
\end{bmatrix} &
\mathbf{Y}_{f,M}^+ &= 
\begin{bmatrix}
{^1}\mathbf{Y}_{f,M}^+ \\
{^2}\mathbf{Y}_{f,M}^+ \\
\vdots \\
{^f}\mathbf{Y}_{f,M}^+
\end{bmatrix} \\
\mathbf{H}_{f}^{ol,u} &=
\begin{bmatrix}
{^1}\mathbf{H}_{f}^{ol,u} \\
{^2}\mathbf{H}_{f}^{ol,u} \\
\vdots \\
{^f}\mathbf{H}_{f}^{ol,u}
\end{bmatrix} &
\mathbf{L}_{f,p} &=
\begin{bmatrix}
{^1}\mathbf{L}_{f,p} \\
{^2}\mathbf{L}_{f,p} \\
\vdots \\
{^f}\mathbf{L}_{f,p}
\end{bmatrix}
\end{align}
\end{subequations}
with ${^i}\mathbf{U}_{f,M}^+ \in \mathbb{R}^{n_u \times M}$, ${^i}\mathbf{Y}_{f,M}^+ \in \mathbb{R}^{n_y \times M}$, ${^i}\mathbf{H}_{f}^{ol,u} \in \mathbb{R}^{n_y \times n_u f}$ and  ${^i}\mathbf{L}_{f,p} \in \mathbb{R}^{n_y \times (n_u + n_y) p}$, $i \in \left\{ 1, f \right\}$. Then, for each $i \in \left\{ 1, f \right\}$,
\begin{equation}\label{eq22:blockdataequ}
{^i}\mathbf{Y}_{f,M}^+ = {^i}\mathbf{L}_{f,p} \mathbf{Z}_{p,M}^- + {^i}\mathbf{H}_{f}^{ol,u} \mathbf{U}_{f,M}^+ + {^i}\mathbf{N}_{f,M}^+ .
\end{equation}
When we look closer at $\mathbf{H}_{f}^{ol,u}$, it is obvious that
\begin{equation}
\begin{bmatrix}
{^1}\mathbf{H}_{f}^{ol,u} \\
{^2}\mathbf{H}_{f}^{ol,u} \\
\vdots \\
{^f}\mathbf{H}_{f}^{ol,u}
\end{bmatrix} =
\begin{bmatrix}
\mathbf{D} & \mathbf{0} & \cdots & \mathbf{0}\\
\mathbf{C}\mathbf{B} & \mathbf{D} & \cdots &\mathbf{0}\\
\vdots & \ddots & \ddots &\vdots\\
\mathbf{C}\mathbf{A}^{f-2}\mathbf{B} & \cdots & \mathbf{C}\mathbf{B} & \mathbf{D}
\end{bmatrix} .
\end{equation}
Thus, the last columns of ${^i}\mathbf{H}_{f}^{ol,u}$ are all composed of zeros except for ${^f}\mathbf{H}_{f}^{ol,u}$. By taking into account this characteristic, Eq.~\eqref{eq22:blockdataequ} becomes\footnote{For $i=1$, ${^i}\bar{\mathbf{H}}_{f}^{ol,u} = \mathbf{D}$.}
\begin{multline}\label{eq22:linregcompcausalmatdataequ}
{^i}\mathbf{Y}_{f,M}^+ = {^i}\mathbf{L}_{f,p} \mathbf{Z}_{p,M}^- \\
+ 
\underbrace{\begin{bmatrix}
\mathbf{C}\mathbf{A}^{i-2}\mathbf{B} & \cdots & \mathbf{C}\mathbf{B} & \mathbf{D}
\end{bmatrix}}_{{^i}\bar{\mathbf{H}}_{f}^{ol,u}}
\begin{bmatrix}
{^1}\mathbf{U}_{f,M}^+ \\
{^2}\mathbf{U}_{f,M}^+ \\
\vdots \\
{^i}\mathbf{U}_{f,M}^+
\end{bmatrix} + {^i}\mathbf{N}_{f,M}^+, \ i \in \left\{ 1, f \right\}
\end{multline}
by removing the terms which involve the zero-block matrices of $\mathbf{H}_{f}^{ol,u}$. The corresponding input signals $\left({^{i+1}}\mathbf{U}_{f,M}^+, \cdots,  {^{f}}\mathbf{U}_{f,M}^+ \right)$ are future signals with respect to ${^i}\mathbf{Y}_{f,M}^+$. Thus, when a fully-parameterized Toeplitz matrix is considered, the matrix data equation~\eqref{eq22:linregcompmatdataequ} contains non-causal terms. The approach based on Eq.~\eqref{eq22:linregcompcausalmatdataequ} excludes these non-causal terms and, in a way, makes the model ``more identifiable''. As previously, the matrices ${^i}\mathbf{L}_{f,p}$ and ${^i}\bar{\mathbf{H}}_{f}^{ol,u}$ can be estimated for $i \in \left\{ 1, f \right\}$ from Eq.~\eqref{eq22:linregcompcausalmatdataequ} as a least-squares problem, the solution of which is given by
\begin{equation}
\begin{bmatrix}
{^i}\hat{\mathbf{L}}_{f,p} \\
{^i}\hat{\bar{\mathbf{H}}}_{f}^{ol,u}
\end{bmatrix} =
{^i}\mathbf{Y}_{f,M}^+
\begin{bmatrix}
 \mathbf{Z}_{p,M}^- \\
{^1}\mathbf{U}_{f,M}^+ \\
{^2}\mathbf{U}_{f,M}^+ \\
\vdots \\
{^i}\mathbf{U}_{f,M}^+
\end{bmatrix}^\dag
\end{equation}
where ${\bullet}^\dag$ stands for the Moore-Penrose pseudo-inverse of the matrix $\bullet$ \cite{GL96}. Finally, by stacking all the estimates ${^i}\hat{\mathbf{L}}_{f,p}$ and ${^i}\hat{\bar{\mathbf{H}}}_{f}^{ol,u}$ for $i \in \left\{ 1, f \right\}$, we get $\hat{\mathbf{L}}_{f,p}$ and $\hat{\mathbf{H}}_{f}^{ol,u}$.

\begin{remark}
All these solutions yield estimates with a smaller asymptotic variance in comparison with the ordinary least-squares estimates obtained in Paragraph~\ref{para22:ordlssol} \cite{Pet95}. 
\end{remark}

\paragraph{Improvements by using a rank constraint}

Up until now, the rank constraint $\rank{\left( \mathbf{L}_{f,p} \right)} = n_x$ has not been taken into account during the estimation procedure. Now, when the aforementioned least-squares algorithms are employed, full rank estimates $\hat{\mathbf{L}}_{f,p}$ are generated. In order to incorporate this prior restriction, a singular value decomposition of $\hat{\mathbf{L}}_{f,p}$ can be used. More precisely,
\begin{equation}\label{eq22:svdLp}
\bm{\mathcal{W}}_l \hat{\mathbf{L}}_{f,p} \bm{\mathcal{W}}_r =
\begin{bmatrix}
\bm{\mathcal{U}}_s & \bm{\mathcal{U}}_n
\end{bmatrix}
\begin{bmatrix}
\bm{\Sigma}_s & \mathbf{0} \\
\mathbf{0} & \bm{\Sigma}_n
\end{bmatrix}
\begin{bmatrix}
\bm{\mathcal{V}}_s^\top \\
\bm{\mathcal{V}}_n^\top
\end{bmatrix}
\end{equation}
where $\bm{\mathcal{W}}_l$ and $\bm{\mathcal{W}}_r$ are weighting matrices chosen by the user in order to allow the construction of various estimates (see \cite{Bau05} for a discussion about this problem). Herein, $\bm{\Sigma}_s$ contains the $n_x$ largest singular values of $\bm{\mathcal{W}}_l \hat{\mathbf{L}}_{f,p} \bm{\mathcal{W}}_r$. Thus, we get
\begin{equation}
\hat{\mathbf{L}}_{f,p} = \bm{\mathcal{U}}_s \bm{\Sigma}_s \bm{\mathcal{V}}_s^\top .
\end{equation}
This SVD combined with the unconstrained regression introduced previously is a reduced rank regression approach \cite{Bau03}.

\begin{remark}
Again, in relation to the comments made in Paragraph~\ref{para22:ordlssol}, the SVD~\eqref{eq22:svdLp} can be related to the SVD of $\mathbf{R}_{32}$ applied in the PO-MOESP algorithm \cite{Ver94} in order to extract the extended observability subspace.
\end{remark}

% SUBSUBSECTION %
\subsubsection{Extraction of the state-space matrices}\label{para22:ssmatextractionol}

In the literature, many solutions are available to extract the state-space matrices from the subspaces estimated beforehand \cite{Bau05,Qin06}. Again, two main basic ideas can be emphasized, the other developments being adapted versions of these main techniques. The first class of techniques aims at constructing a state sequence from $\hat{\mathbf{L}}_{f,p}$ and the past data. The second one tries to extract the observability subspace from $\hat{\mathbf{L}}_{f,p}$. Both are quickly described in the following.

% Paragraph
\paragraph{State subspace approach}\label{para22:statessmatextractionol}

As shown in Sub-Section~\ref{para22:stateconstruct}, the state sequence can be related to past input and past output signals with the help of $\bm{\Omega}_p(\tilde{\mathbf{A}},\begin{bmatrix} \bm{\mathcal{K}} & \tilde{\mathbf{B}} \end{bmatrix})$ as follows
\begin{equation}
\bar{\mathbf{x}}(t) = 
\begin{bmatrix} 
\bm{\Omega}_p(\tilde{\mathbf{A}},\bm{\mathcal{K}}) & \bm{\Omega}_p(\tilde{\mathbf{A}},\tilde{\mathbf{B}}) 
\end{bmatrix}
\begin{bmatrix}
\mathbf{y}_{p}^-(t) \\
\mathbf{u}_{p}^-(t)
\end{bmatrix} =
\bm{\Omega}_p(\tilde{\mathbf{A}},\begin{bmatrix} \bm{\mathcal{K}} & \tilde{\mathbf{B}} \end{bmatrix})
\mathbf{z}_{p}^-(t) .
\end{equation}
By getting back to the definition of $\mathbf{L}_{f,p}$, it is obvious that a rank $n_x$ estimate of $\bm{\Omega}_p(\tilde{\mathbf{A}},\begin{bmatrix} \bm{\mathcal{K}} & \tilde{\mathbf{B}} \end{bmatrix})$ can be obtained from the SVD of $\mathbf{L}_{f,p}$ (see Eq.~\eqref{eq22:svdLp}). Indeed,
\begin{equation}
\hat{\bm{\Omega}}_p(\tilde{\mathbf{A}},\begin{bmatrix} \bm{\mathcal{K}} & \tilde{\mathbf{B}} \end{bmatrix}) =
\bm{\Sigma}_s^{1/2} \bm{\mathcal{V}}_s^\top \bm{\mathcal{W}}_r^{-1} .
\end{equation}
Thus, from $\hat{\bm{\Omega}}_p(\tilde{\mathbf{A}},\begin{bmatrix} \bm{\mathcal{K}} & \tilde{\mathbf{B}} \end{bmatrix})$, we can approximate the state sequence as follows
\begin{equation}
\hat{\mathbf{x}}(t) = \hat{\bm{\Omega}}_p(\tilde{\mathbf{A}},\begin{bmatrix} \bm{\mathcal{K}} & \tilde{\mathbf{B}} \end{bmatrix}) \mathbf{z}_{p}^-(t) .
\end{equation}
This procedure can be adapted to get an accurate estimate of $\hat{\mathbf{x}}(t+1)$ by using a shifted estimated state sequence \cite{Bau03}. Alternatives have been suggested, \emph{e.g.}, in \cite{OM96} or more recently in \cite{CP04c}. For instance, the approach developed by {P. Van Overschee} and {B. De Moor}, which is based on specific non-steady state Kalman filters, relies on two different initial state vectors for $\hat{\mathbf{x}}(t)$ and $\hat{\mathbf{x}}(t+1)$ leading to unbiased estimates. See \cite{OM94b} and the discussion available in \cite[Chapter 4]{OM96} for details. Now, by having access to $\hat{\mathbf{x}}(t)$ as well as $\hat{\mathbf{x}}(t+1)$, the state-space matrices can be estimated in one step from the least-squares fitting
\begin{equation}\label{eq22:lsestABCD}
\begin{bmatrix}
\hat{\mathbf{A}} & \hat{\mathbf{B}} \\
\hat{\mathbf{C}} & \hat{\mathbf{D}}
\end{bmatrix} =
\arg \min_{\mathbf{A},\mathbf{B},\mathbf{C},\mathbf{D}} 
\sum_{k=1}^N 
\left\| 
\begin{bmatrix}
\hat{\mathbf{x}}(k+1) \\
\mathbf{y}(k)
\end{bmatrix} -
\begin{bmatrix}
\mathbf{A} & \mathbf{B} \\
\mathbf{C} & \mathbf{D}
\end{bmatrix}
\begin{bmatrix}
\hat{\mathbf{x}}(k) \\
\mathbf{u}(k)
\end{bmatrix}
\right\|^2_2 .
\end{equation}
This least-squares estimation leads to consistent parameters when the length of the data set used for this linear regression tends to infinity. A two-step procedure is also available in the literature (see, \emph{e.g.}, \cite{LM96}).% According to the author's experience, most of these techniques lead to equivalent results.

% Paragraph
\paragraph{Observability subspace approach}\label{para22:observsubspaceextraction}

By using the SVD~\eqref{eq22:svdLp} differently, the observability subspace $\range{\left( \bm{\Gamma}_{f}(\mathbf{A},\mathbf{C}) \right)}$ can be recovered. Indeed,
\begin{equation}
\hat{\bm{\Gamma}}_{f}(\mathbf{A},\mathbf{C}) = \bm{\mathcal{W}}_l^{-1} \bm{\mathcal{U}}_s \bm{\Sigma}_s^{1/2} .
\end{equation}
Then, from $\hat{\bm{\Gamma}}_{f}(\mathbf{A},\mathbf{C})$, the matrices $\mathbf{A}$ and $\mathbf{C}$ can be extracted as follows
\begin{subequations}
\begin{align}
\hat{\mathbf{C}} &= \hat{\bm{\Gamma}}_{f}(\mathbf{A},\mathbf{C}) [1:n_y,:] \\
\hat{\mathbf{A}} &= \hat{\bm{\Gamma}}_{f}(\mathbf{A},\mathbf{C}) [n_y+1:end,:] \left(\hat{\bm{\Gamma}}_{f}(\mathbf{A},\mathbf{C}) [1:(f-1) n_y,:]\right)^\dag .
\end{align}
\end{subequations}
Finally, the matrices $\mathbf{B}$ and $\mathbf{D}$ can be estimated from the linear regression \cite{VV07}
\begin{equation}
\mathbf{y}(t) =
\begin{bmatrix}
\sum_{\tau = 0}^{t-1} \mathbf{u}^\top(\tau) \otimes \hat{\mathbf{C}} \hat{\mathbf{A}}^{t-\tau-1} & \mathbf{u}^\top(t) \otimes \mathbf{I}_{n_y \times n_y}
\end{bmatrix}
\begin{bmatrix}
\vect{\left( \mathbf{B} \right)} \\
\vect{\left( \mathbf{D} \right)}
\end{bmatrix} .
\end{equation}
The aforementioned solutions can be viewed as the standard techniques used to extract the state-space matrices $(\mathbf{A},\mathbf{B},\mathbf{C},\mathbf{D})$ from an estimate of the observability subspace. Many alternatives are suggested in the literature \cite{Ver94,OM96,Lov97,DSC04}. An interesting guideline is also available in \cite{CP01}.

\begin{remark}
  Again, according to the way the state-matrices $(\mathbf{A},\mathbf{B},\mathbf{C},\mathbf{D})$ are estimated, many solutions can be put forward for the estimation of the Kalman gain $\bm{\mathcal{K}}$. Basically, two main families can be pointed out. The first one resorts to dedicated Riccati equations and can be related to the stochastic realization theory. The second one relies on a residual technique, \emph{i.e.}, the Kalman gain $\bm{\mathcal{K}}$ is estimated from the variance of the residuals resulting from the least-squares problem~\eqref{eq22:lsestABCD}. The interested reader can consult \cite{LM96,CP01,Bau03,Bau05,Qin06} and the references therein for the most popular solutions. See also \cite[Chapter 9]{VV07} for a method which leads to a guaranteed stabilizing estimate of $\bm{\mathcal{K}}$.
\end{remark}

%%%%%%%%%%%%%%%%%%%%%%%%%%% SUBSECTION %%%%%%%%%%%%%%%%%%%%%%%%%%
\subsection[Subspace ident. using linear regression for closed-loop systems]{Subspace-based identification using linear regression for systems operating in closed-loop}\label{para22:ls4sidclosedloop} 

\begin{figure}[htbp]
  \centering
\ifx\JPicScale\undefined\def\JPicScale{1}\fi
\psset{unit=\JPicScale mm}
\psset{linewidth=0.3,dotsep=1,hatchwidth=0.3,hatchsep=1.5,shadowsize=1,dimen=middle}
\psset{dotsize=0.7 2.5,dotscale=1 1,fillcolor=black}
\psset{arrowsize=1 2,arrowlength=1,arrowinset=0.25,tbarsize=0.7 5,bracketlength=0.15,rbracketlength=0.15}
\begin{pspicture}(0,0)(120,32)
\pspolygon[](29,20)(45,20)(45,10)(29,10)
\rput(37,15){$\bm{\mathcal{C}}$}
\psline{->}(45,15)(55,15)
\rput{0}(60,15){\psellipse[](0,0)(5,5)}
\psline(57,18)(63,12)
\psline(57,12)(63,18)
\psline{->}(60,30)(60,20)
\psline{->}(65,15)(75,15)
\rput{0}(14,15){\psellipse[](0,0)(5,5)}
\psline(11,18)(17,12)
\psline(11,12)(17,18)
\psline{->}(19,15)(29,15)
\pspolygon[](75,20)(91,20)(91,10)(75,10)
\rput(83,15){$\bm{\mathcal{S}}$}
\psline{->}(91,15)(101,15)
\rput{0}(106,15){\psellipse[](0,0)(5,5)}
\psline(103,18)(109,12)
\psline(103,12)(109,18)
\psline{->}(106,30)(106,20)
\psline{->}(111,15)(120,15)
\psline(116,15)(116,0)
\psline(116,0)(14,0)
\psline{->}(14,0)(14,10)
\rput(14,12){-}
\rput(57,15){+}
\rput(60,18){+}
\rput(106,18){+}
\rput(103,15){+}
\psline{->}(83,30)(83,20)
\rput(60,32){$r_2(t)$}
\rput(83,32){$w(t)$}
\rput(106,32){$v(t)$}
\rput(69,18){$u(t)$}
\rput(116,18){$y(t)$}
\psline{->}(0,15)(9,15)
\rput(4,18){$r_1(t)$}
\rput(11,15){+}
\end{pspicture}
  \caption{Block diagram of a system $\bm{\mathcal{S}}$ operating in closed-loop with a controller $\bm{\mathcal{C}}$.}
  \label{fig22:closedloopscheme}
\end{figure}
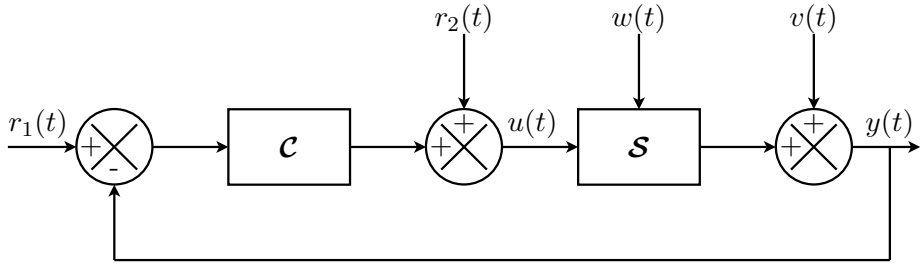

In many situations, the data set used to identify the process must be collected under closed-loop conditions (see Fig.~\ref{fig22:closedloopscheme}). Such an experimental procedure can be required for safety reasons (an unstable plant that requires control for instance) but also in order to maintain the quality of the production during the identification. Notice also that, in identification for control \cite{HSH09}, interesting results (sometimes better than under open-loop conditions) can be obtained when the system works in closed-loop (linearization of the behavior of the plant, reduction of the tests duration, less restrictive excitation conditions, ...) \cite{Gev97,Gev06,Bal06b,Bal08}.

As far as subspace-based identification for closed-loop systems is concerned, most of the standard algorithms have problems and give biased estimates when the data is collected in closed-loop (see \cite[Lemma 1]{LM96} for a proof). The problem inherent to the standard subspace-based identification algorithms under closed-loop conditions can be highlighted by considering again Eq.~\eqref{eq22:linregcompvectordataequ}, \emph{i.e.}, the vector data equation
\begin{equation}\label{eq22:linregcompvectordataequ2}
\mathbf{y}_{f}^+(t) = \mathbf{L}_{f,p} \mathbf{z}_{p}^-(t) + \mathbf{H}_{f}^{ol,u} \mathbf{u}_{f}^+(t) +  \mathbf{n}_{f}^+(t) .
\end{equation}
When the data is collected in open-loop, the noise term $\mathbf{n}_{f}^+$ and the future stacked input vector $\mathbf{u}_{f}^+$ are uncorrelated. Thus, in the open-loop framework, this least-squares-based problem can lead to unbiased estimates as shown previously in Sub-Section~\ref{para22:ls4sidopenloop}. When the system works in closed-loop, this property is no more satisfied because the feedback introduces a correlation between the input and the noise. Thus, biased estimates are obtained when open-loop MOESP \cite{VD92,VD92b,Ver93}, CVA \cite{Lar90} or N4SID \cite{OM94b} algorithms are used with closed-loop data. The first attempts devoted to the closed-loop subspace-based identification have consisted in extending the MOESP or N4SID algorithms 
\begin{itemize}
\item by modifying the instrumental variable used to decrease the noise effect \cite{Ver93b,CV97},
\item by resorting to the prior knowledge available on the controller in order to adapt the open-loop N4SID algorithm with closed-loop data \cite{OM97}.
\end{itemize}
Interesting from a practical point of view, the contributions till the middle of the 1990's can be viewed as extensions of the standard open-loop methods. Recent developments have improved the performance of the subspace-based identification algorithms under closed-loop conditions \cite{LM96,Jan03,QL03,Jan05,CP05,QLL05,Chi07,Chi10}. Most of these techniques share the same basic idea and can perform similarly with closed-loop as well as open-loop data. These techniques can be classified as members of the direct approach class (see \cite[Chapter 11]{Kat05} for a definition of the standard classification of the closed-loop methods). Indeed, they practically ignore the existence of the feedback loop and try to estimate the transfer of the plant directly from the signals $\mathbf{u}$ and $\mathbf{y}$ (see Fig.~\ref{fig22:closedloopscheme}). As highlighted by the studies available, \emph{e.g.}, in \cite{LQL05,Chi06,Qin06}, these techniques
\begin{itemize}
\item use high-order ARX models (HOARX) at least in one step of the procedure,
\item rewrite or modify the vector data equation~\eqref{eq22:linregcompvectordataequ2} in order to uncorrelate the input signals and the noise term.
\end{itemize}
Three of them (the innovation estimation method (IEM) \cite{QL03}, the state-space ARX (SSARX) technique \cite{Jan03} and the prediction-based subspace identification (PBSID) \cite{Chi07} are introduced in the following of this sub-section. They can be considered as the best recent contributions for subspace-based closed-loop system identification and are at the heart of the main developments concerning subspace-based identification for LTI systems till the 2000's \cite{Qin06,Chi07b,Chi10}. 

\begin{remark}
In the sequel, we will mainly focus on the PBSIDopt algorithm, \emph{i.e.}, the optimized version of the PBSID algorithm \cite{Chi07b}. However, because most of the theoretical developments and properties verified by this optimized algorithm have been initially proved for the un-optimized prediction-based subspace identification algorithm or the Whitening Filter Algorithm (WFA) \cite{CP05}, the acronym PBSID will be mainly used in the following. These algorithms mainly differ from the way they are implemented. They are indeed asymptotically equivalent \cite{Chi07,Chi07b,Chi10}.
\end{remark}

Because it is considered in the following that the system works in closed-loop, three conditions commonly used for the direct closed-loop state-space system identification are adopted hereafter.
\begin{assumption}\label{ass:excitationclosedloop}
  The external excitation\footnote{Most of the time, the signal $\mathbf{r}_1$ is fixed equal to zero for simplification.} $\mathbf{r} = \mathbf{r}_2 + \bm{\mathcal{C}}(q^{-1}) \mathbf{r}_1$ (see Fig.~\ref{fig22:closedloopscheme}) is a zero-mean sequence 
  \begin{itemize}
  \item uncorrelated with the process noise $\mathbf{w}$ and the measurement noise $\mathbf{v}$,
\item sufficiently persistently exciting \cite{SS89,Lju99}.
  \end{itemize}  
\end{assumption}
Up until now, no assumption has been made concerning the correlation between the input $\mathbf{u}$ and the noise sequences $\mathbf{v}$ and $\mathbf{w}$. The good excitation property of the external excitation $\mathbf{r}$ ensures that the input sequence $\mathbf{u}$ has sufficient excitation in order to excite correctly the dynamics of the system to identify (see \cite{SS89,Lju99,Wal05,Bal06b} for a general discussion concerning the percistency of excitation).
\begin{assumption}\label{ass:delay}
The feedback loop contains at least one sample delay.
\end{assumption}
Physically, Assumption~\ref{ass:delay} implies that the system or the controller has no direct feed-through. Theoretically, Assumption~\ref{ass:delay} ensures the identifiability of the transfer function of the plant $\bm{\mathcal{S}}$ (see \cite{GA81,GA82} for a discussion about this property and the following consequences). Furthermore, it guarantees that the innovation sequence\footnote{Remember that the innovation sequence is a stationary zero-mean white noise process.} $\mathbf{e}(j)$ and the input $\mathbf{u}(k)$ are uncorrelated $\forall \ j \geq k$ \cite{LQL04}.
\begin{assumption}\label{ass:closedloopstab}
The closed-loop system (see Fig.~\ref{fig22:closedloopscheme}) is asymptotically stable.
\end{assumption}

% It is important to point out that, in the following of this sub-section, we are concerned with algorithms which aim at estimating the open-loop state-space matrices $(\mathbf{A}, \mathbf{B}, \mathbf{C}, \mathbf{D})$ from the available input and output samples $\left\{\mathbf{u}(t)\right\}^N_{t=1}$ and $\left\{\mathbf{y}(t)\right\}^N_{t=1}$. Thus, no model for the feedback channel is looked for. This is in contrast with, \emph{e.g.}, the so-called joint input-output identification techniques \cite{SS89} where models of the open-loop system as well as the controller are estimated. Notice also that the identification algorithms introduced hereafter can be directly applied to data acquired on a system operating in open-loop.

% SUBSECTION %
\subsubsection{The innovation estimation method}

As said previously, the main problem with closed-loop data is the correlation of the future inputs with the past output measurements or the past noise. Indeed, this correlation makes the traditional subspace-based identification methods biased. In order to bypass this difficulty, the innovation estimation method (IEM) \cite{QL03} iteratively pre-estimates the past innovation sequence. Then, by using this estimate, the observability subspace can be extracted.

The starting point of the IEM is (again) the matrix data equation
\begin{equation}\label{eq22:linregcompmatdataequ2}
\mathbf{Y}_{f,M}^+ = \mathbf{L}_{f,p} \mathbf{Z}_{p,M}^- + \mathbf{H}_{f}^{ol,u} \mathbf{U}_{f,M}^+ + \mathbf{N}_{f,M}^+
\end{equation}
where (see Sub-Section~\ref{para22:unconstlssol})
\begin{equation}
\mathbf{N}_{f,M}^+ =  \mathbf{H}_{f}(\mathbf{A}, \bm{\mathcal{K}}, \mathbf{C}, \mathbf{I}_{n_y}) \mathbf{E}_{f,M}^+ 
\end{equation}
and
\begin{equation}
 \mathbf{H}_{f}(\mathbf{A}, \bm{\mathcal{K}}, \mathbf{C}, \mathbf{I}_{n_y}) = 
\begin{bmatrix} 
\mathbf{I}_{n_y \times n_y} & \mathbf{0} & \cdots & \mathbf{0}\\
\mathbf{C} \bm{\mathcal{K}} & \mathbf{I}_{n_y \times n_y} & \cdots &\mathbf{0}\\
\vdots & \ddots & \ddots &\vdots\\
\mathbf{C} \mathbf{A}^{f-2} \bm{\mathcal{K}} & \cdots & \mathbf{C} \bm{\mathcal{K}} & \mathbf{I}_{n_y \times n_y}
\end{bmatrix} .
\end{equation}
By resorting again to the row-block partitioning of~\eqref{eq22:linregcompmatdataequ2} introduced in Paragraph~\ref{para22:improvedcalcHf}, we get
\begin{multline}\label{eq22:partitionmatrices}
{^i}\mathbf{Y}_{f,M}^+ = \mathbf{C} \mathbf{A}^{i-1} 
\bm{\Omega}_p(\tilde{\mathbf{A}},\begin{bmatrix} \bm{\mathcal{K}} & \tilde{\mathbf{B}} \end{bmatrix}) \mathbf{Z}_{p,M}^- \\ + 
\begin{bmatrix}
\mathbf{C} \mathbf{A}^{i-2} \mathbf{B} & \cdots & \mathbf{C} \mathbf{B}
\end{bmatrix}
\begin{bmatrix}
{^1}\mathbf{U}_{f,M}^+ \\
\vdots \\
{^{i-1}}\mathbf{U}_{f,M}^+
\end{bmatrix}
+ \mathbf{D} \ {^i}\mathbf{U}_{f,M}^+ \\ +
\begin{bmatrix}
\mathbf{C} \mathbf{A}^{i-2} \bm{\mathcal{K}} & \cdots & \mathbf{C} \bm{\mathcal{K}} & \mathbf{I}_{n_y \times n_y}
\end{bmatrix}
\begin{bmatrix}
{^1}\mathbf{E}_{f,M}^+ \\
\vdots \\
{^i}\mathbf{E}_{f,M}^+
\end{bmatrix}, \ i \in \left\{ 1, f \right\} .
\end{multline}
By looking closer at this equation, it is interesting to stress that the innovation matrix $\mathbf{E}_{f,M}^+$ can be partitioned into two parts (corresponding to the past innovation $\begin{bmatrix} {{^1}\mathbf{E}_{f,M}^+}^\top & \cdots & {{^{i-1}}\mathbf{E}_{f,M}^+}^\top \end{bmatrix}^\top$ and the future innovation ${^i}\mathbf{E}_{f,M}^+$ respectively) thanks to the structure of the matrix $\mathbf{H}_{f}(\mathbf{A}, \bm{\mathcal{K}}, \mathbf{C}, \mathbf{I}_{n_y})$, \emph{i.e.},
\begin{multline}
\begin{bmatrix}
\mathbf{C} \mathbf{A}^{i-2} \bm{\mathcal{K}} & \cdots & \mathbf{C} \bm{\mathcal{K}} & \mathbf{I}_{n_y \times n_y}
\end{bmatrix}
\begin{bmatrix}
{^1}\mathbf{E}_{f,M}^+ \\
\vdots \\
{^i}\mathbf{E}_{f,M}^+
\end{bmatrix} = \\
\begin{bmatrix}
\mathbf{C} \mathbf{A}^{i-2} \bm{\mathcal{K}} & \cdots & \mathbf{C} \bm{\mathcal{K}}
\end{bmatrix}
\begin{bmatrix}
{^1}\mathbf{E}_{f,M}^+ \\
\vdots \\
{^{i-1}}\mathbf{E}_{f,M}^+
\end{bmatrix} +
{^i}\mathbf{E}_{f,M}^+ , \ i \in \left\{ 1, f \right\} .
\end{multline}
Thus, when $\mathbf{D} = \mathbf{0}$, by having access to an estimate of ${^k}\mathbf{E}_{f,M}^+$, $k < i$, an unbiased estimate of $\mathbf{C} \mathbf{A}^{i-1} \bm{\Omega}_p(\tilde{\mathbf{A}},\begin{bmatrix} \bm{\mathcal{K}} & \tilde{\mathbf{B}} \end{bmatrix})$ can be calculated from closed-loop data through a straightforward linear regression because the future innovation ${^i}\mathbf{E}_{f,M}^+$ is uncorrelated with $ \mathbf{Z}_{p,M}^-$, ${^k}\mathbf{E}_{f,M}^+$ and ${^k}\mathbf{U}_{f,M}^+$, $k < i$, even under closed-loop conditions.

As said previously, this procedure requires the availability of an accurate estimate of ${^k}\mathbf{E}_{f,M}^+$, $k < i$. To get this estimate, a multi-stage least-squares algorithm is suggested in \cite{QL03}. The starting point is Eq.~\eqref{eq22:partitionmatrices} for $i=1$ with $\mathbf{D} = \mathbf{0}$, \emph{i.e.},
\begin{equation}
{^1}\mathbf{Y}_{f,M}^+ = \mathbf{C} 
\bm{\Omega}_p(\tilde{\mathbf{A}},\begin{bmatrix} \bm{\mathcal{K}} & \tilde{\mathbf{B}} \end{bmatrix}) \mathbf{Z}_{p,M}^- + {^1}\mathbf{E}_{f,M}^+ 
\end{equation}
which is a VARX\footnote{A VARX model is similar to a high-order ARX model when SISO systems are handled.} (Vector Auto-Regressive with eXogenous inputs) model. In spite of the feedback, ${^1}\mathbf{E}_{f,M}^+$ is uncorrelated with $\mathbf{Z}_{p,M}^-$. Thus, an unbiased estimate of $\mathbf{C} \bm{\Omega}_p(\tilde{\mathbf{A}},\begin{bmatrix} \bm{\mathcal{K}} & \tilde{\mathbf{B}} \end{bmatrix})$ can be obtained from closed-loop data as a least-squares estimate
\begin{equation}\label{eq22:hoarxiem}
\widehat{\mathbf{C} \bm{\Omega}}_p(\tilde{\mathbf{A}},\begin{bmatrix} \bm{\mathcal{K}} & \tilde{\mathbf{B}} \end{bmatrix}) = {^1}\mathbf{Y}_{f,M}^+ {\mathbf{Z}_{p,M}^-}^\dag
\end{equation}
and a least-squares estimate of the innovation process is
\begin{equation}
{^1}\hat{\mathbf{E}}_{f,M}^+ = \mathbf{Y}_{f,M}^+ - \widehat{\mathbf{C} \bm{\Omega}}_p(\tilde{\mathbf{A}},\begin{bmatrix} \bm{\mathcal{K}} & \tilde{\mathbf{B}} \end{bmatrix}) \mathbf{Z}_{p,M}^- .
\end{equation}
Then, by knowing ${^1}\hat{\mathbf{E}}_{f,M}^+$, unbiased estimates of $\mathbf{C} \mathbf{A} \bm{\Omega}_p(\tilde{\mathbf{A}},\begin{bmatrix} \bm{\mathcal{K}} & \tilde{\mathbf{B}} \end{bmatrix})$ and ${^2}\mathbf{E}_{f,M}^+$ can be calculated from Eq.~\eqref{eq22:partitionmatrices} for $i=2$. This procedure is repeated iteratively for $i=1$ to $f$. Finally, from the least-squares estimates of $\mathbf{C} \bm{\Omega}_p(\tilde{\mathbf{A}},\begin{bmatrix} \bm{\mathcal{K}} & \tilde{\mathbf{B}} \end{bmatrix})$, $\mathbf{C} \mathbf{A} \bm{\Omega}_p(\tilde{\mathbf{A}},\begin{bmatrix} \bm{\mathcal{K}} & \tilde{\mathbf{B}} \end{bmatrix})$, ..., $\mathbf{C} \mathbf{A}^{f-1} \bm{\Omega}_p(\tilde{\mathbf{A}},\begin{bmatrix} \bm{\mathcal{K}} & \tilde{\mathbf{B}} \end{bmatrix})$, we can reconstruct
\begin{equation}
\begin{bmatrix}
\mathbf{C} \\
\mathbf{C} \mathbf{A} \\
\vdots \\
\mathbf{C} \mathbf{A}^{f-1}
\end{bmatrix} \bm{\Omega}_p(\tilde{\mathbf{A}},\begin{bmatrix} \bm{\mathcal{K}} & \tilde{\mathbf{B}} \end{bmatrix}) =
\bm{\Gamma}_f(\mathbf{A},\mathbf{C}) \bm{\Omega}_p(\tilde{\mathbf{A}},\begin{bmatrix} \bm{\mathcal{K}} & \tilde{\mathbf{B}} \end{bmatrix}) .
\end{equation}
As a final step, the observability subspace can be extracted through a weighted singular value decomposition (see Paragraph~\ref{para22:constleastsquaressol} for details). By knowing an accurate estimate of $\bm{\Gamma}_f(\mathbf{A},\mathbf{C})$, the state-space matrices can be obtained by following the techniques described in Paragraph~\ref{para22:observsubspaceextraction} .

% SUBSUBSECTION %
\subsubsection{Closed-loop subspace-based identification with Markov parameter pre-estimation}

Although the previous technique is developed to deal with closed-loop data, the IEM can have numerical problems when unstable systems are handled \cite{LQL05,CP05}. This drawback is mainly due to the direct use of the state matrix $\mathbf{A}$ in Eq.~\eqref{eq22:linregcompmatdataequ2} and, more problematic when unstable systems are involved, $\mathbf{A}^i$. In order to circumvent this difficulty, {M. Jansson} \cite{Jan03,Jan05}, then {A. Chiuso} and {G. Picci} \cite{CP05,CP05b}, suggested using a predictor form instead of an innovation form of the system. More precisely, we consider
\begin{subequations} \label{eq22:sspredform2}
\begin{align}
\mathbf{x}(t+1) &= \tilde{\mathbf{A}} \mathbf{x}(t) + \tilde{\mathbf{B}} \mathbf{u}(t) + \bm{\mathcal{K}} \mathbf{y}(t) \\
\mathbf{y}(t) &= \mathbf{C} \mathbf{x}(t) + \mathbf{D} \mathbf{u}(t) + \mathbf{e}(t) \label{eq22:outputpredform2}
\end{align}
\end{subequations}
where
\begin{subequations}
\begin{align}
\tilde{\mathbf{A}} &= \mathbf{A} - \bm{\mathcal{K}} \mathbf{C} \\
\tilde{\mathbf{B}} &= \mathbf{B} - \bm{\mathcal{K}} \mathbf{D} .
\end{align}
\end{subequations}
Indeed, as shown previously, one of the advantages of this form is that $\tilde{\mathbf{A}}$ can be constrained to be (exponentially) stable by fixing the gain $\bm{\mathcal{K}}$ correctly so that the eigenvalues of $\tilde{\mathbf{A}}$ are as close to zero as possible (see, again, Assumption~\ref{ass:tildeAnilpotent})

By iterating the equations composing Eq.~\eqref{eq22:sspredform2}, we get the vector data equation
\begin{multline}\label{eq22:vectordataequpredform2}
  \mathbf{y}_{f}^+(t) =\bm{\Gamma}_f(\tilde{\mathbf{A}},\mathbf{C}) \mathbf{x}(t) + \underbrace{\mathbf{H}_{f}(\tilde{\mathbf{A}},\tilde{\mathbf{B}},\mathbf{C},\mathbf{D})}_{\mathbf{H}_{f}^{cl,u}} \mathbf{u}_{f}^+(t) \\ + \underbrace{\mathbf{H}_{f}(\tilde{\mathbf{A}},\bm{\mathcal{K}},\mathbf{C},\mathbf{0})}_{\mathbf{H}_{f}^{cl,y}} \mathbf{y}_{f}^+(t) + \mathbf{e}_{f}^+(t) .
\end{multline}
% where
% \begin{equation}
% \bm{\Gamma}_f(\tilde{\mathbf{A}},\mathbf{C}) =
% \begin{bmatrix}
% \mathbf{C}^\top & \cdots & (\mathbf{C} \tilde{\mathbf{A}}^{f-1})^\top
% \end{bmatrix}^\top \in \mathbb{R}^{n_y f \times n_x} 
% \end{equation}
% \begin{equation}\label{eq22:tildeHf}
% \tilde{\mathbf{H}}_f = 
% \begin{bmatrix} 
% \mathbf{D}& \mathbf{0} & \cdots & \mathbf{0}\\
% \mathbf{C} \tilde{\mathbf{B}} & \mathbf{D} & \cdots &\mathbf{0}\\
% \vdots & \ddots & \ddots &\vdots\\
% \mathbf{C} \tilde{\mathbf{A}}^{f-2} \mathbf{B} & \cdots & \mathbf{C}\mathbf{B} & \tilde{\mathbf{D}}
% \end{bmatrix} \in \mathbb{R}^{n_y f \times n_u f} 
% \end{equation}
% \begin{equation}\label{eq22:tildeGf}
% \tilde{\mathbf{G}}_f = 
% \begin{bmatrix} 
% \mathbf{0}_{n_y} & \mathbf{0} & \cdots & \mathbf{0}\\
% \mathbf{C} \bm{\mathcal{K}} & \mathbf{0}_{n_y} & \cdots &\mathbf{0}\\
% \vdots & \ddots & \ddots &\vdots\\
% \mathbf{C} \tilde{\mathbf{A}}^{f-2} \bm{\mathcal{K}} & \cdots & \mathbf{C} \bm{\mathcal{K}} & \mathbf{0}_{n_y}
% \end{bmatrix} \in \mathbb{R}^{n_y f \times n_y f} .
% \end{equation}
From the same set of equations, we know that
\begin{equation}
\mathbf{x}(t) = \tilde{\mathbf{A}}^p \mathbf{x}(t-p) + \bm{\Omega}_p(\tilde{\mathbf{A}},\begin{bmatrix} \bm{\mathcal{K}} & \tilde{\mathbf{B}} \end{bmatrix}) \mathbf{z}_{p}^-(t) .
\end{equation}
% where
% \begin{subequations}
% \begin{align}
% \tilde{\bm{\Omega}}_p &=
% \begin{bmatrix}
% \tilde{\mathbf{B}} & \tilde{\mathbf{A}} \tilde{\mathbf{B}} & \cdots & \tilde{\mathbf{A}}^{p-1} \tilde{\mathbf{B}} 
% \end{bmatrix} \\
% \tilde{\bm{\Delta}}_p &=
% \begin{bmatrix}
% \bm{\mathcal{K}} & \tilde{\mathbf{A}} \bm{\mathcal{K}} & \cdots & \tilde{\mathbf{A}}^{p-1} \bm{\mathcal{K}}
% \end{bmatrix} .
% \end{align}
% \end{subequations}
Then, with Assumption~\ref{ass:tildeAnilpotent}, similarly to the open-loop case, the contribution of $\mathbf{x}(t-p)$ can be neglected and the vector data equation~\eqref{eq22:vectordataequpredform2} becomes
\begin{multline}
\mathbf{y}_{f}^+(t) = \bm{\Gamma}_{f} (\tilde{\mathbf{A}},\mathbf{C}) \bm{\Omega}_p(\tilde{\mathbf{A}},\begin{bmatrix} \bm{\mathcal{K}} & \tilde{\mathbf{B}} \end{bmatrix}) \mathbf{z}_{p}^-(t)\\ + \mathbf{H}_{f}^{cl,u} \mathbf{u}_{f}^+(t) + \mathbf{H}_{f}^{cl,y} \mathbf{y}_{f}^+(t) + \mathbf{e}_{f}^+(t)
\end{multline}
with standard notations. Under closed-loop conditions, the future innovation $\mathbf{e}_{f}^+$ is correlated with the future stacked vectors $\mathbf{y}_{f}^+$ and $\mathbf{u}_{f}^+$. In order to circumvent this difficulty, M. Jansson (who followed the work of {K. Peternell} on the CCA algorithm \cite{Pet95}) was actually the first to suggest estimating the Toeplitz matrices $\mathbf{H}_{f}^{cl,u}$ and $\mathbf{H}_{f}^{cl,y}$ in a first phase, then using these estimates in order to circumvent this correlation problem \cite{Jan03}. This basic idea is also shared by the PBSID algorithm developed by {A. Chiuso} \cite{Chi07} (see Paragraph~\ref{para22:pbsid}). The similarity between these two algorithms is not restricted to this common step. Indeed, as proved in \cite{Chi06,Chi07}, the SSARX algorithm and the PBSID one (as well as its initial version named the Whitening Filter Algorithm (WFA) \cite{CP05}) are asymptotically equivalent \cite{Chi06,Chi07}.

In both cases, specific matrices involved in the estimation problem are built from estimates of $\mathbf{D}$, $\mathbf{C} \tilde{\mathbf{A}}^{k} \mathbf{B}$ and $\mathbf{C} \tilde{\mathbf{A}}^{k} \bm{\mathcal{K}}$, $k=0, \ \cdots, \ f-2$. These Markov parameters can be calculated from the one-step-ahead VARX model defined as \cite{Chi07}
\begin{equation}\label{eq22:onesteppredictor}
\mathbf{y}(t|t-1) =
\sum_{i=0}^\ell {\tilde{\bm{\mathcal{M}}}_{\mathbf{u}(t-i)} \mathbf{u}(t-i)} + \sum_{i=1}^\ell {\tilde{\bm{\mathcal{M}}}_{\mathbf{y}(t-i)} \mathbf{y}(t-i)}
\end{equation}
where $\ell$  is a user-defined truncation index and $\mathbf{y}(t|t-1)$ is the predicted output at time instant $t$ which uses the inputs from $t$ to $t-\ell$ and the outputs from $t-1$ to $t-\ell$. Again, this finite order long ARX model causes misspecifications and bias when the index $\ell$ is too small. Raising its value a lot can lead to a prohibitive increase of the variance of the estimated matrices. Thus, a trade-off between bias and variance is necessary. It is interesting to point out that, when Assumption~\ref{ass:tildeAnilpotent} is verified, the truncation index can be chosen equal to $p$ which leads to a null truncation error. Furthermore, with this assumption, the relation between this predictor and the Markov parameters is straightforward by substituting Eq.~\eqref{eq22:stateapprox} for $\mathbf{x}$ in Eq.~\eqref{eq22:outputpredform2}. More precisely \cite{CP05},
\begin{subequations}
\begin{align}
\tilde{\bm{\mathcal{M}}}_{\mathbf{u}(t-i)} &=
\left\{
\begin{matrix}
\mathbf{D} & \text{ if } i=0 \\
\mathbf{C} \tilde{\mathbf{A}}^{i-1} \tilde{\mathbf{B}} & \text{ if } i>0
\end{matrix}
\right. \\
\tilde{\bm{\mathcal{M}}}_{\mathbf{y}(t-i)} &=\mathbf{C} \tilde{\mathbf{A}}^{i-1} \bm{\mathcal{K}} .
\end{align}
\end{subequations}
As proved in \cite[Theorem 4.3]{Chi07}, when Assumptions~\ref{ass:noise}, \ref{ass:minimal}, \ref{ass:tildeAnilpotent}, \ref{ass:excitationclosedloop} and \ref{ass:delay} are satisfied, consistent estimates of the Markov parameters of the system can be provided from the predictor~\eqref{eq22:onesteppredictor} even with closed-loop data. Again, like in the open-loop framework, Assumption~\ref{ass:tildeAnilpotent} is necessary to ensure that the truncation error term omitted in Eq.~\eqref{eq22:onesteppredictor} is equal to zero.

\begin{remark}
When a sufficiently large amount of data is available, the estimation of the aforementioned Markov parameters can be performed by minimizing the following least-squares cost function
\begin{equation}
  \left\| {^1}\mathbf{Y}_{f,M}^+  - \begin{bmatrix}
\tilde{\bm{\mathcal{M}}}_{\mathbf{y}(t-\ell)} & \cdots & \tilde{\bm{\mathcal{M}}}_{\mathbf{u}(t-1)} & \tilde{\bm{\mathcal{M}}}_{\mathbf{u}(t)}
\end{bmatrix}
\begin{bmatrix}
\mathbf{Y}_{\ell,M}^- \\
\mathbf{U}_{\ell,M}^- \\
{^1}\mathbf{U}_{f,M}^+
\end{bmatrix} \right\|_F^2
\end{equation}
or, written differently,
\begin{equation}\label{eq22:costfuniem}
  \left\| {^1}\mathbf{Y}_{f,M}^+  - \begin{bmatrix}
\mathbf{C} \tilde{\mathbf{A}}^{\ell-1} \begin{bmatrix} \bm{\mathcal{K}} & \tilde{\mathbf{B}}
\end{bmatrix} & \cdots &
\mathbf{C} \begin{bmatrix} \bm{\mathcal{K}} & \tilde{\mathbf{B}} 
\end{bmatrix} & \mathbf{D}
\end{bmatrix}
\begin{bmatrix}
\mathbf{Z}_{\ell,M}^- \\
{^1}\mathbf{U}_{f,M}^+
\end{bmatrix} \right\|_F^2
\end{equation}
where the involved regressors are defined in Paragraph~\ref{para22:improvedcalcHf}. This cost function is, up to the value of the index $\ell$, the one used in the first step of the IEM algorithm (see Eq.~\eqref{eq22:hoarxiem}). Thus, basically, apart from the algorithmic implementation, the main difference between the IEM and the PDSID-like algorithms only rests on the use of $\bm{\Gamma}_{f}(\mathbf{A},\mathbf{C})$ instead of $\bm{\Gamma}_{f}(\tilde{\mathbf{A}},\mathbf{C})$.
\end{remark}

By having access to accurate estimates of $\mathbf{D}$, $\mathbf{C} \tilde{\mathbf{A}}^{k} \mathbf{B}$ and $\mathbf{C} \tilde{\mathbf{A}}^{k} \bm{\mathcal{K}}$, $k=0, \ \cdots, \ f-2$, the SSARX and the PBSID algorithms mainly\footnote{As claimed by {A. Chiuso} in \cite{Chi07b}, the PBSID algorithm can be viewed as ``a geometrical version of the SSARX algorithm''.} differ from the way these estimates are used in the following steps. See also \cite[Remark 4.5]{Chi07} for a thorough discussion about the main differences between the PBSIDopt and the SSARX algorithms.

\begin{remark}
The reader can see that this basic idea is inspired by the work of {K. Peternell} \cite{Pet95} with his CCA algorithms \cite{PSD96}. The link with the VARX one-step ahead predictor model suggested in \cite{LM96} is also obvious.
\end{remark}

% Paragraph
\paragraph{The state-space {ARX} algorithm}\label{para22:ssarx}

From the estimates of $\mathbf{C} \tilde{\mathbf{A}}^{k} \mathbf{B}$, $\mathbf{C} \tilde{\mathbf{A}}^{k} \bm{\mathcal{K}}$ and $\mathbf{D}$, $k=0, \ \cdots, \ f-2$, estimates of $\mathbf{H}_{f}^{cl,u}$ and $\mathbf{H}_{f}^{cl,y}$ can be formed. Then, it can be written that
\begin{multline}
\underbrace{\mathbf{y}_{f}^+(t) - \hat{\mathbf{H}}_{f}^{cl,u} \mathbf{u}_{f}^+(t) - \hat{\mathbf{H}}_{f}^{cl,y} \mathbf{y}_{f}^+(t)}_{\mathbf{s}_{f}^+(t)} = \\ \bm{\Gamma}_{f}(\tilde{\mathbf{A}},\mathbf{C}) \bm{\Omega}_p(\tilde{\mathbf{A}},\begin{bmatrix} \bm{\mathcal{K}} & \tilde{\mathbf{B}} \end{bmatrix}) \mathbf{z}_{p}^-(t) + \mathbf{e}_{f}^+(t) .
\end{multline}
Because $\mathbf{e}_{f}^+$ and $\mathbf{z}_{p}^-$ are uncorrelated, this equation can be viewed as a low rank linear regression problem for the estimation of $\bm{\Gamma}_{f}(\tilde{\mathbf{A}},\mathbf{C}) \bm{\Omega}_p(\tilde{\mathbf{A}},\begin{bmatrix} \bm{\mathcal{K}} & \tilde{\mathbf{B}} \end{bmatrix})$. Many algorithms are available to perform this optimization. For instance, a least-squares minimization can be carried out in order to get an estimate of $\bm{\Gamma}_{f}(\tilde{\mathbf{A}},\mathbf{C}) \bm{\Omega}_p(\tilde{\mathbf{A}},\begin{bmatrix} \bm{\mathcal{K}} & \tilde{\mathbf{B}} \end{bmatrix})$. Then, a (weighted) SVD can be used in order to extract the extended observability subspace. Instead of resorting to such a least-squares approach, M. Jansson suggests performing a canonical correlation analysis (CCA) on $\mathbf{s}_{f}^+(t)$ and $\mathbf{z}_{p}^-(t)$. More precisely, the following SVD is used\footnote{The aforementioned sample correlation matrices are defined as $\hat{\mathbf{R}}_{\mathbf{q} \mathbf{g}}(t) = \frac{1}{N} \sum_{t=1}^N \mathbf{q}(t) \mathbf{g}^\top(t)$ .}
\begin{equation}
\hat{\mathbf{R}}_{\mathbf{s}_{f}^- \mathbf{s}_{f}^-}^{-1/2} \hat{\mathbf{R}}_{\mathbf{s}_{f}^- \mathbf{z}_{p}^-} \hat{\mathbf{R}}_{\mathbf{z}_{p}^- \mathbf{z}_{p}^-}^{-1/2} = 
\begin{bmatrix}
\bm{\mathcal{U}}_s & \bm{\mathcal{U}}_n
\end{bmatrix}
\begin{bmatrix}
\bm{\Sigma}_s & \mathbf{0} \\
\mathbf{0} & \bm{\Sigma}_n
\end{bmatrix}
\begin{bmatrix}
\bm{\mathcal{V}}_s^\top \\
\bm{\mathcal{V}}_n^\top
\end{bmatrix}
\end{equation}
and the CCA estimate of $\bm{\Omega}_p(\tilde{\mathbf{A}},\begin{bmatrix} \bm{\mathcal{K}} & \tilde{\mathbf{B}} \end{bmatrix})$ is given by
\begin{equation}
\hat{\bm{\Omega}}_p(\tilde{\mathbf{A}},\begin{bmatrix} \bm{\mathcal{K}} & \tilde{\mathbf{B}} \end{bmatrix}) =  \bm{\mathcal{V}}_s^\top \hat{\mathbf{R}}_{\mathbf{z}_{p}^- \mathbf{z}_{p}^-}^{1/2} .
\end{equation}
By having access to $\hat{\bm{\Omega}}_p(\tilde{\mathbf{A}},\begin{bmatrix} \bm{\mathcal{K}} & \tilde{\mathbf{B}} \end{bmatrix})$, the estimated state sequence satisfies
\begin{equation}
\hat{\mathbf{x}}(t) = \hat{\bm{\Omega}}_p(\tilde{\mathbf{A}},\begin{bmatrix} \bm{\mathcal{K}} & \tilde{\mathbf{B}} \end{bmatrix}) \mathbf{z}_{p}^-(t)
\end{equation}
from which the state-space matrices of the system can be estimated by a linear regression as it was introduced in Paragraph~\ref{para22:statessmatextractionol}.

\begin{remark}
A slightly different implementation of the SSARX algorithm is put forward in \cite{Jan05}. A three step procedure is more precisely considered, involving the idea of the ``one step correction method'' of \cite{WJ04}. See \cite{Jan05} for details.
\end{remark}

% Paragraph
\paragraph{The prediction-based subspace identification algorithm}\label{para22:pbsid}

Instead of resorting to the estimates of $\tilde{\mathbf{G}}_{f}$ and $\tilde{\mathbf{H}}_{f}$, the key idea of the prediction-based subspace identification (PBSID) algorithm consists in observing that the product of the state and the extended observability matrix is given by
\begin{equation}\label{eq22:Gammaclx}
\bm{\Gamma}_{f}(\tilde{\mathbf{A}},\mathbf{C}) \mathbf{x}(t) = \bm{\Gamma}_{f}(\tilde{\mathbf{A}},\mathbf{C}) \bm{\Omega}_p(\tilde{\mathbf{A}},\begin{bmatrix} \bm{\mathcal{K}} & \tilde{\mathbf{B}} \end{bmatrix}) \mathbf{z}_{p}^-(t) .
\end{equation}
If this product can be estimated, the extended observability subspace as well as the state sequence can be extracted by solving a low rank optimization problem. To do so, an estimate of $\bm{\Gamma}_{f}(\tilde{\mathbf{A}},\mathbf{C}) \bm{\Omega}_p(\tilde{\mathbf{A}},\begin{bmatrix} \bm{\mathcal{K}} & \tilde{\mathbf{B}} \end{bmatrix})$ must be available. As shown by {A. Chiuso} in \cite{Chi07}, this block-matrix can be constructed from the Markov parameters obtained from the least-squares solution of the optimization problem~\eqref{eq22:costfuniem}. Indeed, it is obvious that
\begin{multline}
\bm{\Gamma}_{f}(\tilde{\mathbf{A}},\mathbf{C}) \bm{\Omega}_p(\tilde{\mathbf{A}},\begin{bmatrix} \bm{\mathcal{K}} & \tilde{\mathbf{B}} \end{bmatrix}) \\ =
\begin{bmatrix}
\mathbf{C} \tilde{\mathbf{A}}^{p-1} \begin{bmatrix} \bm{\mathcal{K}} & \tilde{\mathbf{B}} \end{bmatrix} & \mathbf{C} \tilde{\mathbf{A}}^{p-2} \begin{bmatrix} \bm{\mathcal{K}} & \tilde{\mathbf{B}} \end{bmatrix} & \cdots & \mathbf{C} \begin{bmatrix} \bm{\mathcal{K}} & \tilde{\mathbf{B}} \end{bmatrix} \\
\mathbf{C} \tilde{\mathbf{A}}^{p} \begin{bmatrix} \bm{\mathcal{K}} & \tilde{\mathbf{B}} \end{bmatrix} & \mathbf{C} \tilde{\mathbf{A}}^{p-1} \begin{bmatrix} \bm{\mathcal{K}} & \tilde{\mathbf{B}} \end{bmatrix} & \ddots & \mathbf{C} \tilde{\mathbf{A}} \begin{bmatrix} \bm{\mathcal{K}} & \tilde{\mathbf{B}} \end{bmatrix} \\
\vdots &  & \ddots & \vdots \\
\mathbf{C} \tilde{\mathbf{A}}^{f+p-2} \begin{bmatrix} \bm{\mathcal{K}} & \tilde{\mathbf{B}} \end{bmatrix} & \cdots & \cdots & \mathbf{C} \tilde{\mathbf{A}}^{f-1} \begin{bmatrix} \bm{\mathcal{K}} & \tilde{\mathbf{B}} \end{bmatrix}
\end{bmatrix} 
\end{multline}
where the block-matrices composing this equation are solutions of the least-squares regression~\eqref{eq22:costfuniem}. Furthermore, under Assumption~\ref{ass:tildeAnilpotent}, the former matrices become upper block triangular matrices, \emph{i.e.},
\begin{multline}
\bm{\Gamma}_{f}(\tilde{\mathbf{A}},\mathbf{C}) \bm{\Omega}_p(\tilde{\mathbf{A}},\begin{bmatrix} \bm{\mathcal{K}} & \tilde{\mathbf{B}} \end{bmatrix}) \\ =
\begin{bmatrix}
\mathbf{C} \tilde{\mathbf{A}}^{p-1} \begin{bmatrix} \bm{\mathcal{K}} & \tilde{\mathbf{B}} \end{bmatrix} & \mathbf{C} \tilde{\mathbf{A}}^{p-2} \begin{bmatrix} \bm{\mathcal{K}} & \tilde{\mathbf{B}} \end{bmatrix} & \cdots & \mathbf{C} \begin{bmatrix} \bm{\mathcal{K}} & \tilde{\mathbf{B}} \end{bmatrix} \\
\mathbf{0} & \mathbf{C} \tilde{\mathbf{A}}^{p-1} \begin{bmatrix} \bm{\mathcal{K}} & \tilde{\mathbf{B}} \end{bmatrix} & \ddots & \mathbf{C} \tilde{\mathbf{A}} \begin{bmatrix} \bm{\mathcal{K}} & \tilde{\mathbf{B}} \end{bmatrix} \\
\vdots &  & \ddots & \vdots \\
\mathbf{0} & \cdots & \mathbf{0} & \mathbf{C} \tilde{\mathbf{A}}^{f-1} \begin{bmatrix} \bm{\mathcal{K}} & \tilde{\mathbf{B}} \end{bmatrix}
\end{bmatrix}
\end{multline}
when $f$ is chosen equal to $p$. By recalling Eq.~\eqref{eq22:Gammaclx}, an estimate of the state sequence can be computed from a dedicated weighted SVD from which the state-space matrices of the system can be estimated.

\begin{remark}
The former description can be viewed as a sketch of the PBSID algorithm, really close to the PBSIDopt algorithm \cite{Chi07}. Different versions of this algorithm are available in the literature. The interested reader can study, \emph{e.g.}, \cite{CP05,Chi07,Chi10} for more details. For instance, a recent study \cite{HWV09b} deals with the PBSID algorithm with the help of a Vector Auto-Regressive Moving Average with eXogenous inputs (VARMAX) model. By using this model (instead of a VARX (see Eq.~\eqref{eq22:onesteppredictor})), Assumption~\ref{ass:tildeAnilpotent} can be satisfied with a lower value of the user-defined parameter $p$ (see Paragraph~\ref{para22:finitestateseqapprox} and the discussion about Assumption~\ref{ass:breveAnilpotent} for details). Many details are now available in \cite{Hou11} and really good and reliable implementations of the corresponding subspace-based identification algorithms can be downloaded from Delft Center for Systems and Control under the generic name PBSIDToolbox \cite{Hou10}. As claimed by the authors of this toolbox, ``the Predictor-Based Subspace Identification Toolbox enables you to perform a batch or recursive identification (in open-loop and closed-loop) of LTI/LPV/Hammerstein/Wiener systems''.
\end{remark}

%%%%%%%%%%%%%%%%%%%%%%%%%%%%%%%%%%%%%%%%%%%%%%%%%%%%%%%%%%%%%%%%%
%%%%%%%%%%%%%%%%%%%%%%%%%%%%% SECTION %%%%%%%%%%%%%%%%%%%%%%%%%%%
\section{Conclusions}

In this paper, an overview of the main regression-based subspace identification techniques has been introduced. More precisely, a specific attention has been paid to their algorithmic structures, their common points and their differences. Thus, it has been shown that, whatever the operating conditions (open-loop or closed-loop), several subspace identification schemes can be derived from similar least-squares problem formulations. All these regression-based techniques share indeed the idea of approximating the state sequence of the system with past input and output data. Basically, they mainly differ from the way this state sequence is technically approximated. By following this general idea, it has been highlighted that the most efficient least-squares subspace-based identification techniques rely on a state sequence approximation derived from a reformulation of the well-known innovation form into the predictor-form. Said differently, it has also been shown that, for open-loop and closed-loop systems, the main and well-known subspace-based identification algorithms can all be interpreted as prediction-based subspace algorithm or, at least, involve a prediction step.

\bibliographystyle{plain}

% \bibliography{H:/Research/Publications/Bibliography}
% \bibliography{/media/DATA/Home/Research/ScientificProductions/Bibliography}
% \bibliography{H:/Research/Publications/Bibliography}
% \bibliography{G:/Research/Publications/Bibliography

\end{document}